\documentclass[10pt, twocolumn, twoside]{IEEEtran}

\IEEEoverridecommandlockouts

\usepackage{cite}
\usepackage{graphicx}
\usepackage[cmex10]{amsmath}
\usepackage{amssymb}
\usepackage{array}
\usepackage{psfrag}
\usepackage{color}

\usepackage{multirow}

\usepackage{footnote}
\makesavenoteenv{tabular}

\newtheorem{theorem}{Theorem}

\newtheorem{remark}{Remark}

\newcommand{\mb}{\mathbf}

\newcommand{\uh}{{\mb{h}}}
\newcommand{\uha}{{\uh_1}}
\newcommand{\uhad}{{\uh_{\text{d},1}}}

\newcommand{\uhar}{{\uh_{\text{r},1}}}

\newcommand{\uhaH}{\uh_1^{\cal{H}}}

\newcommand{\uhda}{{\mb{h}_{\text{d},1}}}

\newcommand{\uhra}{{\mb{h}_{\text{r},1}}}

\newcommand{\umu}{\pmb{\mu}}
\newcommand{\umuH}{{\pmb{\mu}}^{\cal{H}}}

\newcommand{\uLambda}{{\pmb{\Lambda}}}

\newcommand{\uSigma}{{\pmb{\Sigma}}}

\newcommand{\unu}{{\pmb{\nu}}}

\newcommand{\ur}{{\mb{r}}}

\newcommand{\uA}{{\mb{A}}}

\newcommand{\uQ}{{\mb{Q}}}

\newcommand{\uH}{{\mb{H}}}
\newcommand{\uHH}{{\mb{H}}^{\cal{H}}}
\newcommand{\uHhat}{\widehat{\uH}}
\newcommand{\uHhatH}{\widehat{\uH}^{\cal{H}}}

\newcommand{\uHb}{{\mb{H}_2}}
\newcommand{\uHbH}{{\mb{H}_2^{\cal{H}}}}

\newcommand{\uHd}{{\uH_{\text{d}}}}

\newcommand{\uHdb}{{\uH_{\text{d},2}}}

\newcommand{\uHdH}{{\uH_{\text{d}}^{\cal{H}}}}

\newcommand{\uHdn}{{\uH_{\text{d,n}}}}

\newcommand{\uHr}{{\uH_{\text{r}}}}

\newcommand{\uHrb}{{\uH_{\text{r},2}}}

\newcommand{\uHrn}{{\uH_{\text{r,n}}}}

\newcommand{\NT}{{N_{\text{T}}}}

\newcommand{\NR}{{N_{\text{R}}}}
\newcommand{\uP}{{\mb{P}}}
\newcommand{\uR}{{\mb{R}}}

\newcommand{\uRT}{{\mathbf{R}_{\text{T}}}}

\newcommand{\uRTinv}{{\mathbf{R}_{\text{T}}^{-1}}}
\newcommand{\uRTK}{{\mathbf{R}_{\text{T,}K}}}

\newcommand{\uv}{{\mb{v}}}

\newcommand{\ud}{{\mb{d}}}
\newcommand{\ux}{{\mb{x}}}

\newcommand{\uxa}{{\mb{x}_1}}
\newcommand{\uxaH}{\mb{x}_1^{\cal{H}}}

\newcommand{\uz}{{\mb{z}}}

\newcommand{\mzero}{\mb{0}}

\newcommand{\mbR}{{\mb{R}}}

\newcommand{\uU}{{\mb{U}}}

\newcommand{\uUH}{{\mb{U}}^{\cal{H}}}
\newcommand{\uV}{{\mb{V}}}
\newcommand{\uVH}{{\mb{V}}^{\cal{H}}}
\newcommand{\uW}{{\mb{W}}}
\newcommand{\uWinv}{{\uW}^{-1}}

\newcommand{\uWhat}{\widehat{\uW}}

\newcommand{\uT}{{\mb{T}}}
\newcommand{\uTH}{{\mb{T}^{\cal{H}}}}
\newcommand{\uTinv}{\mb{T}^{-1}}
\newcommand{\uTinvH}{(\mb{T}^{-1})^{\cal{H}}}

\newcommand{\mbI}{\mb{I}}

\newcommand{\RT}{\mathbf{R}_{\text{T}}}

\newcommand{\RTinvaa}{\left[\mathbf{R}_{\text{T},K}^{-1}\right]_{1,1}}

\newcommand{\RTaloneinvaa}{\left[\mathbf{R}_{\text{T}}^{-1}\right]_{1,1}}
\newcommand{\RTaa}{\mathbf{R}_{{\text{T},K}_{11}}}

\newcommand{\RTbb}{\mathbf{R}_{{\text{T},K}_{22}}}

\newcommand{\RTbbinva}{\mathbf{R}_{{\text{T},K}_{22}}^{-1}}

\newcommand{\RTab}{\mathbf{R}_{{\text{T},K}_{12}}}

\newcommand{\RTba}{\mathbf{R}_{{\text{T},K}_{21}}}
\newcommand{\rTba}{\mathbf{r}_{{\text{T},K}_{21}}}
\newcommand{\rTbaH}{\mathbf{r}_{{\text{T},K}_{21}}^{\cal{H}}}

\title{Exact MIMO Zero-Forcing Detection Analysis for Transmit-Correlated Rician Fading}

\author{
Constantin Siriteanu\thanks{C.~Siriteanu and A.~Takemura are with the Department of Mathematical Informatics, Graduate School of Information Science and Technology, University of Tokyo, Japan, and Japan Science and Technology Agency, CREST.},
Steven D. Blostein\thanks{S.~D.~Blostein and S.~Yousefi are with the Department of Electrical and Computer Engineering, Queen's University, Canada.}, Akimichi Takemura, \\
Hyundong Shin\thanks{H.~Shin is with the Department of Electronics and Radio Engineering, Kyung Hee University, South Korea.},
Shahram Yousefi, Satoshi Kuriki\thanks{S.~Kuriki is with the Institute of Statistical Mathematics, Tokyo,
Japan.}}

\begin{document}


\maketitle

\begin{abstract}
We analyze the performance of multiple input/mul-tiple output (MIMO) communications systems employing spatial multiplexing and zero-forcing detection (ZF).
The distribution of the ZF signal-to-noise ratio (SNR) is characterized when either the intended stream or interfering streams experience Rician fading, and when the fading may be correlated on the transmit side.
Previously, exact ZF analysis based on a well-known SNR expression has been hindered by the noncentrality of the Wishart distribution involved.
In addition, approximation with a central-Wishart distribution has not proved consistently accurate.
In contrast, the following exact ZF study proceeds from a lesser-known SNR expression that separates the intended and interfering channel-gain vectors.
By first conditioning on, and then averaging over the interference, the ZF SNR distribution for Rician--Rayleigh fading is shown to be an infinite linear combination of gamma distributions.
On the other hand, for Rayleigh--Rician fading, the ZF SNR is shown to be gamma-distributed.
Based on the SNR distribution, we derive new series expressions for the ZF average error probability, outage probability, and ergodic capacity.
Numerical results confirm the accuracy of our new expressions, and reveal effects of interference and channel statistics on performance.
\end{abstract}


\begin{IEEEkeywords}
Azimuth spread, $ K $-factor, gamma distribution, MIMO, Rayleigh and Rician (Ricean) fading, transmit correlation, Wishart distribution, zero-forcing.
\end{IEEEkeywords}

\section{Introduction}
\label{section_introduction}

\subsection{Background, Motivation, and Scope}

Multiple input/multiple output (MIMO) wireless communication theory, simulation, and implementation have demonstrated that substantial performance gains are possible by suitable processing at the transmit and receive antennas\cite{marzetta_tit_99}\cite{paulraj_pieee_04}\cite{simon_alouini_book_00}\cite{gesbert_spm_07}\cite{lee_09_jwcn}\cite{nishimori_vtc_09}.
MIMO spatial multiplexing, whereby streams of symbols are transmitted from each antenna, can enhance data and user capacity\cite{gesbert_spm_07}.
However, the effects on MIMO multiplexing performance of fading with nonzero mean and correlation are not yet fully understood even for  low-complexity detection methods such as zero-forcing detection (ZF), which cancels the interference but may enhance the noise, or minimum mean-square error detection (MMSE), also known as optimum combining\cite{louie_tcom_09}, which balances interference and noise but requires knowledge of the noise variance\cite{mckay_tcomm_09}\cite{jiang_tit_11}.

Such knowledge gaps need to be filled because state-of-the-art channel modeling, e.g., WINNER\cite{winner_d_1_1_2_v_1_2}, has revealed that, in most scenarios, measured channels are characterized by nonzero mean, i.e., Rician fading.
Its mean and correlation are determined by the $ K $-factor and the azimuth spread (AS), respectively.
WINNER has characterized measured $ K $ and AS with scenario-dependent lognormal distributions\cite[Table~I]{siriteanu_tvt_11}.
Whereas MMSE has been analyzed exactly for Rician fading in a few publications\cite{louie_tcom_09}\cite{mckay_tcomm_09}\cite{jiang_tit_11}\cite{kim_twc_08},
ZF has so far been analyzed exactly only for Rayleigh fading\cite{winters_tcom_94}\cite{gore_cl_02}\cite{kiessling_spawc_03}\cite{jiang_tit_11}.

Attempting to study ZF for Rician fading by viewing ZF as a limit case of MMSE or by approximating Rician fading with Rayleigh fading may not yield reliable results.
On the one hand, MMSE analysis for Rician fading can be very involved\cite{mckay_tcomm_09}; also, although popular in earlier work\cite[p.~210]{paulraj_pieee_04}, the assertion that MMSE reduces to ZF for vanishing noise has been revised recently by Jiang \textit{et al.} in\cite{jiang_tit_11}.
On the other hand, Siriteanu \textit{et al.}\cite{siriteanu_tvt_11} have found that an approximation of Rician fading with Rayleigh fading (based on approximating a noncentral-Wishart distribution with a central-Wishart distribution of equal mean) is not consistently accurate: accuracy degrades with higher rank of the channel-matrix mean and depends on $ K $ and AS (even for low rank).


Therefore, herein, we develop an exact ZF analysis for:
\begin{enumerate}
\item Rician--Rayleigh fading, i.e., the intended stream undergoes Rician fading, whereas the interfering streams all undergo Rayleigh fading.
\item Rayleigh--Rician fading.
\end{enumerate}
These fading assumptions have previously also been made in analyses of optimum combining and maximal-ratio combining in\cite{louie_tcom_09}\cite{mckay_tcomm_09}\cite{chayawan_tcomm_02}\cite{kang_tcom_02}, where they were justified as relevant to propagation in macrocells and microcells. They are also relevant in heterogeneous networks, e.g., for femtocells deployed within macrocells, as shown in\cite[Fig.~1]{saquib_wc_12}.
Finally, our ZF analysis herein assumes zero receive-correlation but allows for nonzero transmit-correlation.

\subsection{Previous Approaches}
\label{section_previous_work}

For transmit-correlated Rayleigh--Rayleigh fading, Gore \textit{et al.}\cite{gore_cl_02} showed that the ZF signal-to-noise ratio (SNR) is gamma-distributed, based on the central-Wishart distribution of the matrix that appears in the ratio-form expression of this SNR\cite[Eq.~(5)]{gore_cl_02}, by writing the ZF SNR as a Schur complement in the central-Wishart distributed matrix\cite[Eq.~(8)]{gore_cl_02}.
Kiessling and Speidel further expressed this Schur complement as a Hermitian form in\cite[Eq.~(7)]{kiessling_spawc_03}.
(This lesser-known expression for the ZF SNR has more recently appeared in\cite[Eq.~(15)]{jiang_tit_11}.)
Conveniently, in the Hermitian form, the random matrix, which accounts for interference fading, is idempotent\footnote{Matrix $ \uA $ is \textsl{idempotent} if $ \uA^2 = \uA $. Its eigenvalue matrix is then idempotent. Thus, the eigenvalues of $ \uA $ are either $ 0 $ or $ 1 $.}.
Thus,\cite{kiessling_spawc_03} readily showed that the ZF SNR is gamma-distributed by averaging the SNR Hermitian-form expression only over the vector, which accounts for intended fading. (See also\cite[Eq.~(16)]{jiang_tit_11}.)

Since Rician fading yields noncentral-Wishart distribution, the approach from\cite{gore_cl_02}\cite{kiessling_spawc_03} alone can only approximately characterize the SNR distribution, after approximating the noncentral-Wishart distribution with a central-Wishart distribution of equal mean --- see\cite{siriteanu_tvt_11} and references therein.
However,\cite{siriteanu_tvt_11} has found, for Rician--Rician fading, that the approximation is not consistently accurate.
(Herein, we show that it is also inaccurate for Rician--Rayleigh fading.)

For MMSE, the stream-SNR can be written in a ratio form\cite[Eq.~(13)]{jiang_tit_11} similar to that for ZF\cite[Eq.~(5)]{gore_cl_02}, but MMSE analysis, including for Rician fading, has typically proceeded directly from an equivalent Hermitian-form expression\cite[p.~439]{simon_alouini_book_00}\cite[Eq.~(2)]{louie_tcom_09}\cite[Eq.~(7)]{kim_twc_08}\cite[Eq.~(3)]{mckay_tcomm_09}\cite[Eq.~(17)]{jiang_tit_11}.
Then, as for ZF in\cite[Eq.~(7)]{kiessling_spawc_03}, the intended and interfering fading contributions are separated into the vector and matrix of the Hermitian form, respectively.
However, for MMSE, the matrix in the Hermitian form is not idempotent. 
Thus, deriving the SNR moment generating function (m.g.f.) requires tedious averaging also over this matrix.
This is illustrated by McKay \textit{et al.} in\cite{mckay_tcomm_09}, where averaging over the matrix proceeds by averaging over its eigenvalues and eigenvectors to yield the complicated SNR m.g.f.~expressions for Rician--Rayleigh and Rayleigh--Rician fading from\cite[Eqs.~(13)-(19),~(35)-(43)]{mckay_tcomm_09}.


\subsection{Our Approach and Contributions}
\label{section_Contributions}
We use the approach from\cite{gore_cl_02}\cite{kiessling_spawc_03} as the first step in our ZF analysis for transmit-correlated Rician--Rayleigh fading.
In the second step, we average only over the eigenvectors of the idempotent matrix in the SNR Hermitian-form, and obtain a new and exact expression for its m.g.f.~in terms of a confluent hypergeometric function, i.e., as an infinite series.
This reveals that the ZF SNR distribution for Rician--Rayleigh fading is an infinite linear combination of gamma distributions.
(A gamma distribution characterizes the ZF SNR for Rayleigh--Rayleigh fading.)
The SNR m.g.f.~is then written in closed-form.
In the third step, we show that a mean--correlation condition that reduces the infinite linear combination of gamma distributions to a single gamma distribution also helps extend our analysis to Rayleigh--Rician fading.
Finally, we derive for ZF new and exact expressions, in the form of infinite series, for important performance measures such as the average error probability (AEP), outage probability, and ergodic capacity. They reveal interesting effects of $ K $, AS, and interference on ZF performance for Rician--Rayleigh and Rayleigh--Rician fading.
For our infinite-series expressions, we only outline herein the convergence proofs and the computation method and issues.
Details appear in our recent work\cite{siriteanu_ausctw_14}.


\subsection{Notation}
\begin{itemize}
\item Scalars, vectors, and matrices are represented with lowercase italics, lowercase boldface, and uppercase boldface, respectively, e.g., $ a $, $ \uh $, and $ \uH $; the zero vectors and matrices of appropriate dimensions are denoted with $ \mzero $; superscripts $ \cdot^{\cal{T}} $ and $ \cdot^{\cal{H}} $ stand for transpose and Hermitian (i.e., complex-conjugate) transpose; $ [\cdot]_{i,j} $ indicates the $i,j$th element of a matrix; $ \| \uH \|^2 = \sum_{i}^{\NR} \sum_{j}^{\NT} | [\uH]_{i,j} |^2 $ is the squared Frobenius norm of $ \NR \times \NT $ matrix $ \uH $; $ i = 1 : N $ stands for the enumeration $ i = 1, \, 2, \, \ldots \, N $; $ \otimes $ stands for the Kronecker product; $ \propto $ stands for `proportional to'.
\item $ \uh \sim {\cal{CN}} (\uh_{\text{d}}, \uR_{\uh} $) indicates that $ \uh $ is a complex-valued circularly-symmetric Gaussian random vector\cite{paulraj_pieee_04}\cite{gallager_08_prep} with mean (i.e., deterministic component) $ \uh_{\text{d}} $ and covariance $ \mbR_{\uh} $; subscripts $ \cdot_{\text{d}} $ and $ \cdot_{\text{r}} $ identify, respectively, the deterministic and random components of a scalar, vector, or matrix; subscript $ \cdot_{\text{n}} $ indicates a normalized variable; $ \mathbb{E} \{ \cdot \} $ denotes statistical average; $ { \,{\buildrel d \over =}\, } $ indicates random variables equal in distribution; $ Gamma (N, \Gamma_1) $ represents the gamma distribution with shape parameter $ N $ and scale parameter $\Gamma_1 $; $ \chi_m^2(\delta) $ denotes the noncentral chi-square distribution with $ m $ degrees of freedom and noncentrality parameter $ \delta $; $ \chi_m^2 $ denotes the central chi-square distribution with $ m $ degrees of freedom; $ Beta(N, M) $ represents the beta distribution with shape parameters $ N $ and $ M $; $ {\cal{F}} (N, M) $ represents the $ {\cal{F}} $-distribution with degrees of freedom $ N $ and $ M $\cite[Chs.~17, 18, 25, 27]{johnson_book_95}.
\item $ M^{(p)}(s) $ stands for the derivative of order $ p $ of $ M(s) $; $ \gamma (\kappa, x) = \int_{0}^{x} t^{\kappa - 1} e^{-t} \mathrm{d} t $, $ \Gamma (\kappa, x) = \int_{x}^{\infty} t^{\kappa - 1} e^{-t} \mathrm{d} t $, and $ \text{E}_1 (x) = \int_{x}^{\infty} t^{-1}e^{-t} \mathrm{d} t, \; x > 0 $ stand for the functions \textsl{incomplete gamma}, \textsl{complementary incomplete gamma}, and \textsl{exponential integral}, respectively\cite[Eqs.~(8.2.1-2), p.~174, Eq.~(6.2.1), p.~150, Eq.~(6.11.1), p.~153]{NIST_book_10}.
\item $ (N)_n $ is the Pochhammer symbol, i.e.,  $ (N)_0 = 1 $ and $ (N)_n = N (N + 1) \ldots (N + n - 1) $, $ \forall n > 1 $\cite[p.~xiv]{NIST_book_10}, and $ \,_1F_1(\cdot; \cdot; \cdot) $ is the confluent hypergeometric function\cite[Eq.~(13.2.2), p.~322]{NIST_book_10}.
\end{itemize}

\subsection{Paper Organization}
Section~\ref{section_system_channel_model} introduces our statistical models for the receiver noise and channel fading.
Sections~\ref{section_performance_analysis} and~\ref{section_Bartlett} characterize the ZF SNR distribution for Rician--Rayleigh fading and Rayleigh--Rician fading, respectively.
Section~\ref{section_Performance_Measure_Expressions} derives performance-measure expressions for ZF, and outlines their convergence proofs and their computational issues.
Finally, Section~\ref{section_Numerical_Results} presents numerical results from analysis and Monte Carlo simulation.

\section{Signal, Noise, and Fading Models}
\label{section_system_channel_model}

We consider an uncoded multiantenna-based wireless communication system over a frequency-flat fading channel.
We assume that there are $ \NT $ and $ \NR $ antenna elements at the transmitter(s) and receiver, respectively, with $ \NT \le \NR $.
Letting $\ux = [x_1 \, x_2 \, \cdots \, x_{\NT}]^{\cal{T}}$ denote the $ \NT \times 1 $ zero-mean transmit-symbol vector with $ \mathbb{E} \{ \ux \ux^{\cal{H}} \} = \mbI_{\NT} $, the $ \NR \times 1 $ vector with the received signals can be represented as\cite[Eq.~(8)]{paulraj_pieee_04}:
\begin{eqnarray}
\label{equation_system}
\ur = \sqrt{\frac{E_{\text{s}}}{\NT}} \, \uH \ux + \uv = \sqrt{\frac{E_{\text{s}}}{\NT}} \, \uha x_1 + \sqrt{\frac{E_{\text{s}}}{\NT}} \,  \sum_{k = 2}^{\NT} \uh_k x_k + \uv .
\end{eqnarray}

Above, $ E_{\text{s}}/\NT $ is the energy transmitted per symbol (i.e., per antenna), so that $ E_{\text{s}} $ is the energy transmitted per channel use.
The additive noise vector $\uv $ is zero-mean, uncorrelated, circularly-symmetric, complex Gaussian with $ \uv \sim {\cal{CN}} (\mzero, N_0 \, \mbI_{\NR})$.
We will also employ its normalized version $ \uv_{\text{n}} = \uv/\sqrt{N_0} \sim {\cal{CN}} (\mzero, \mbI_{\NR}) $.
In analysis, we will employ the per-symbol input SNR, which is defined as follows:
\begin{eqnarray}
\label{equation_SNR_per_symbol}
 { \Gamma_{\text{s}} }  = \frac{E_{\text{s}}}{N_0} \frac{1}{\NT}.
\end{eqnarray}
On the other hand, in numerical results, we will employ the per-bit input SNR, which, for a modulation constellation with $ M $ symbols (e.g., $ M $PSK),  is defined as follows:
\begin{eqnarray}
\label{equation_SNR_per_bit}
 { \Gamma_{\text{b}} }  = \frac{\Gamma_{\text{s}} }{\log_2 M} = \frac{E_{\text{s}}}{N_0} \frac{1}{\NT} \frac{1}{\log_2 M}.
\end{eqnarray}

Then, $\uH = (\uha \; \uh_2 \; \ldots \; \uh_{\NT}) $ is the $\NR \times \NT $ complex-Gaussian channel matrix, assumed to have rank $ \NT $.
Vector $ \uh_k $ comprises the channel factors between transmit-antenna $ k $ and all receive-antennas.
The deterministic (i.e., mean) and random components of $\uH$ are denoted as $\uHd = (\uhda \; \uh_{\text{d},2} \; \ldots \; \uh_{\text{d},\NT}) $ and $\uHr = (\uhra \; \uh_{\text{r},2} \; \ldots \; \uh_{\text{r},\NT})$, respectively, so that $ \uH = \uHd + \uHr $.
If $ [\uHd]_{i,j} = 0 $ then $ |\left[ \uH \right]_{i,j}| $ has a Rayleigh distribution; otherwise, $ |\left[ \uH \right]_{i,j}| $ has a Rician distribution\cite{simon_alouini_book_00}.
Typically, the channel matrix for Rician fading is written as
\begin{eqnarray}
\label{equation channelH}
\uH =  \uHd + \uHr = \sqrt{\frac{K}{K+1}} \, \uHdn + \sqrt{\frac{1}{K+1}} \, \uHrn,
\end{eqnarray}
where it is assumed for normalization purposes\cite{loyka_twc_09} that $ \| \uHdn \|^2 = \NT \NR $ and $ \mathbb{E} \{ |\left[ \uHrn \right]_{i,j}|^2 \} = 1, \forall i, j $,
so that $ \mathbb{E} \{ \| \uH \|^2 \} = \NT \NR $.
Power ratio
\begin{eqnarray}
\label{equation K_definition}
\frac{\| \uHd \|^2}{\mathbb{E} \{ \| \uHr \|^2 \} } = \frac{ \frac{K}{K+1} \| \uHdn \|^2}{\frac{1}{K+1} \mathbb{E} \{ \| \uHrn \|^2 \} } = K
\end{eqnarray}
is the Rician $ K $-factor: $ K = 0 $ yields Rayleigh fading for all elements of $ \uH $; $ K \neq 0 $ yields Rician fading if $ \uHdn \neq \mzero $.

Finally, we assume zero receive-correlation but allow for nonzero transmit-correlation.
We also need to assume, for tractability, as in previous work\cite{gore_cl_02}\cite{kiessling_spawc_03}, that all conjugate-transposed rows of $ \uHrn $ have distribution $ {\cal{CN}} (\mzero, \uRT ) $, with $ [\uRT]_{i,i} = 1 $, $ \forall i = 1 : \NT$.
Thus, all conjugate-transposed  rows of $ \uHr $ have distribution $ {\cal{CN}} (\mzero, \uRTK = \frac{1}{K + 1 } \uRT  ) $.
The elements of $ \uRT $ can be computed from the AS as shown in\cite[Section~VI.A]{siriteanu_tvt_11} for WINNER, i.e., Laplacian, power azimuth spectrum\footnote{WINNER has also modeled measured  $ K $ (in dB) and AS (in degrees) as random variables with scenario-dependent lognormal distributions.}\cite{winner_d_1_1_2_v_1_2}.

\section{MIMO ZF Performance Analysis for Rician--Rayleigh Fading}
\label{section_performance_analysis}

\subsection{ZF SNR in Conventional (Ratio) Form}
\label{section_SNR_intro}
Given $\uH$, ZF for the signal from~(\ref{equation_system}) means separately mapping each element of the following vector into the closest modulation constellation symbol\cite[Eq.~(22)]{paulraj_pieee_04}:
\begin{eqnarray}
\label{equation ZF_for_perfect_CSI}
\sqrt{\frac{ \NT }{ E_{\text{s}} }} \left[ \uH^{\cal{H}} \uH \right]^{-1} \uH^{\cal{H}} \, \ur
= \ux +\frac{ 1 }{  \sqrt{ { \Gamma_{\text{s}} }  }} \left[ \uH^{\cal{H}} \uH \right]^{-1} \uH^{\cal{H}} \uv_{\text{n}}.
\end{eqnarray}
There is no interference among the transmitted streams, which explains the ZF name for this technique.
The noise vector that corrupts the transmitted signal vector $ \ux $ in~(\ref{equation ZF_for_perfect_CSI}) has correlation matrix $ \frac{ 1 }{  { \Gamma_{\text{s}} } }  \uW^{-1} $, where $ \uW = \uH^{\cal{H}} \uH $.
Therefore, the ZF SNR for stream $ k = 1 $, for instance, has the well-known ratio form
\begin{eqnarray}
\label{equation_gammak_perfect_CSI}
\gamma_1 = \frac{  { \Gamma_{\text{s}} }  }{\left[\uW^{-1}\right]_{1,1}}.
\end{eqnarray}

This form was employed by the original ZF analysis for Rayleigh--Rayleigh fading in\cite{gore_cl_02}, as follows.
First, scalar $ 1/\left[\uW^{-1}\right]_{1,1} $ was written as the determinant of the Schur complement of submatrix $ [\uW]_{2:\NT,2:\NT} $ in $ \uW $\cite[Eq.~(8)]{gore_cl_02}.
(Since this Schur complement is scalar, taking its determinant is unnecessary.)
Then, because $ \uW $ is central-Wishart-distributed\cite{siriteanu_tvt_11}, the Schur complement is chi-square distributed\cite[Eq.~(10)]{gore_cl_02}, i.e., $ \gamma_1 $ is gamma distributed.

\subsection{ZF SNR in Hermitian Form\cite{kiessling_spawc_03}}
\label{section_SNR_intro_Hermitian_form}

When channel-matrix elements are Rician-fading, $ \uW $ has a noncentral-Wishart distribution\cite{siriteanu_tvt_11}. Its complexity has precluded finding the distribution of $ \gamma_1 $ with the approach from\cite{gore_cl_02}, i.e., based on~(\ref{equation_gammak_perfect_CSI}).
We show below that the first step in making this derivation tractable for Rician--Rayleigh fading is rewriting the ZF SNR as a Hermitian form\cite{kiessling_spawc_03}\cite{jiang_tit_11}.

Thus, let us partition the channel matrix $ \uH $ itself\footnote{I.e., instead of partitioning $ \uW = \uH^{\cal{H}} \uH $, as done in\cite{gore_cl_02}.} as
\begin{eqnarray}
\label{equation_partitioned_H}
\uH = \left( \uha \; \uHb \right) = \left( \uhad \; \uHdb \right) + \left( \uhar \; \uHrb \right),
\end{eqnarray}
where $ \uha $ is the  $ \NR \times 1 $ channel vector corresponding to the intended stream, and $ \uHb = ( \uh_2 \; \ldots \; \uh_{\NT} ) $ is the $ \NR \times (\NT - 1) $ matrix with the channel vectors corresponding to the interfering streams.
Then, as in\cite{kiessling_spawc_03}\cite{jiang_tit_11}, we can rewrite $ \gamma_1 $ from~(\ref{equation_gammak_perfect_CSI}) as the Hermitian form\footnote{Note that $ \uhaH \uQ_2 \uha $ is the Schur complement of $ \uHbH \uHb $ in $ \uW $\cite{siriteanu_tit_13}.}
\begin{eqnarray}
\label{equation_gamma1_partitioned_H_xyx}
\label{equation_gamma1_partitioned_H}
\gamma_1 =  { \Gamma_{\text{s}} }  \, \uhaH \underbrace{[  \mbI_{\NR} - \uHb \left( \uHbH \uHb \right)^{-1} \uHbH  ]}_{=\uQ_2} \uha = { \Gamma_{\text{s}} }  \, \uhaH \uQ_2 \uha,
\end{eqnarray}
where $ \NR \times \NR $ Hermitian matrices $ \uP_2 = \uHb \left( \uHbH \uHb \right)^{-1} \uHbH $ and $ \uQ_2 $ are idempotent, have ranks $ \NT - 1 $ and $ N = \NR - \NT + 1 $, respectively, and have eigenvalues as listed below:
\begin{eqnarray}
\label{equation_eigenvalues_Htilde}
\uP_2: & 1, \; 1, \; \ldots, \; 1, & 0, \; 0, \; \ldots, \; 0.
\label{equation_eigenvalues_Q}\\
\uQ_2: & \underbrace{0, \; 0, \; \ldots, \; 0,}_{\NT-1} & \underbrace{1, \; 1, \; \ldots, \; 1}_{N}.
\end{eqnarray}
The m.g.f.~of $ \gamma_1 $ is derived by first conditioning on $ \uHb $ (i.e., $ \uQ_2 $) and then by averaging over it.

\subsection{Expressing $ \gamma_1 $ Conditioned on $ \uHb$ (i.e., $ \uQ_2 $)\cite{kiessling_spawc_03}}
\label{section_Appendix_gamma1_Q}

Let us partition the $ \NT \times \NT $ transmit-covariance matrix $ \uRTK $ according to~(\ref{equation_partitioned_H}), as follows:
\begin{eqnarray}
\label{equation_partitioned_RTK}
\uRTK = \left(
  \begin{array}{cc}
    \RTaa & \RTab \\
    \RTba & \RTbb \\
  \end{array}
  \right) = \left(
  \begin{array}{cc}
     \frac{1}{K + 1}  & \rTbaH \\
    \rTba & \RTbb \\
  \end{array}
  \right).
\end{eqnarray}
The $ (\NT - 1) \times (\NT - 1) $ matrix $ \RTbb $ is the column-covariance matrix of $ \uHb $.
On the other hand, the $ (\NT - 1) \times 1 $ vector $ \rTba $ represents the cross-covariances of elements of columns in $ \uHb $ and corresponding elements of $
\uha $.
Since $ \uha $ and $ \uHb $ in~(\ref{equation_partitioned_H}) are jointly Gaussian, the distribution of $ \uha $ given $ \uHb $ is as follows\cite[Appendix]{kiessling_spawc_03}
\begin{eqnarray}
\label{equation_h1_given_Htilde_1}
\uha | \uHb  \sim {\cal{CN}} \left(\umu + \uHb \ur_{2,1}, \frac{1}{\RTinvaa} \, \mbI_{\NR} \! \right),
\end{eqnarray}
where
\begin{eqnarray}
\label{equation_umu}
\umu & = & \uhda - \uHdb \ur_{2,1}, \\
\label{equation_r21}
\ur_{2,1} & = & \RTbbinva \rTba
\end{eqnarray}
are, respectively, $ \NR \times 1 $ and $ (\NT - 1) \times 1 $ deterministic vectors,
and\footnote{Note that $ \RTaa - \RTab \, \RTbbinva \, \RTba  $ is the Schur complement of $ \RTbb $ in $ \uRTK $, and the variance of $ [\uH]_{1,1} $ given $ [\uH]_{1, 2: \NT} $\cite{siriteanu_tit_13}.}
\begin{eqnarray}
\label{equation_RTKinv11}
\left( \RTinvaa \right)^{-1} & =& \RTaa - \RTab \, \RTbbinva \, \RTba \nonumber \\
& = & \frac{1}{K + 1} - \rTbaH \, \RTbbinva \, \rTba.
\end{eqnarray}
%

Substituting~(\ref{equation_h1_given_Htilde_1}) into~(\ref{equation_gamma1_partitioned_H}) and further manipulating as in\cite{kiessling_spawc_03} helps write the SNR conditioned on $ \uQ_2 $ as
\begin{eqnarray}
\label{equation_gamma1_Kiess_reduction}
\gamma_1 | \uQ_2  & = &  \Gamma_1 \uxaH \uQ_2 \uxa,
\end{eqnarray}
where
\begin{eqnarray}
\label{equation_Gamma}
\Gamma_1 & = & \frac{ { \Gamma_{\text{s}} } }{\RTinvaa} \propto \frac{ { \Gamma_{\text{s}} } }{K+1}, \\
\label{equation_x1}
\uxa & \sim & {\cal{CN}} \left( {\sqrt{\RTinvaa}} \, \umu, \mbI_{\NR} \right).
\end{eqnarray}
Notice that $ \gamma_1  $ depends on $ \ur_{2,1} $ (i.e., $ \uRT $) through $ \umu = \uhad - \uHdb \ur_{2,1} $.
However, the assumption $ \uHdb = \mzero $ made later, in Section~\ref{section_special_case}, yields $ \umu = \uhad $, which removes this dependence.
Then, $ \uRT $ will affect the ZF SNR $ \gamma_1 $ only through scalar $ \RTinvaa $.

\subsection{Derivation of the M.G.F. of the Conditioned SNR}
\label{section_Conditional_mgf_SNR}

Using Turin's result from\cite[Eq.~(4a)]{turin_bio_60}, the m.g.f.~of $ \gamma_1|\uQ_2 $ from~(\ref{equation_gamma1_Kiess_reduction}) can be written as
\begin{eqnarray}
\label{equation_gamma1_mgf_Turin}
M_{\gamma_1|\uQ_2}(s) = \mathbb{E}_{{\gamma_1} | \uQ_2} \{ e^{s \gamma_1} | \uQ_2 \} \quad \quad\quad \quad \quad\quad\quad \quad\quad\quad \; \;\nonumber \\ = \frac{\exp \big\{ - \RTinvaa  \umuH \left[ \mbI_{\NR} - \left( \mbI_{\NR} - s  \Gamma_1 \uQ_2 \right)^{-1} \right] \umu \big\}}{ \det \left( \mbI_{\NR} - s \Gamma_1 \uQ_2 \right)}.
\end{eqnarray}
The natural next step is to average $ M_{\gamma_1|\uQ_2}(s) $ from~(\ref{equation_gamma1_mgf_Turin}) over $ \uQ_2 $, which is done in the next subsection.
However, this averaging requires further manipulation of $ M_{\gamma_1|\uQ_2}(s) $.

First, let us consider the singular value decomposition $ \uHb = \uU \uSigma \uVH $, where $ \NR \times \NR $ matrix $ \uU $ and $ (\NT - 1) \times (\NT - 1 )$ matrix $ \uV $ are unitary, i.e., $ \uUH \uU = \uU \uUH = \mbI_{\NR} $ and $ \uVH \uV = \uV \uVH = \mbI_{\NT-1} $, and $ \NR \times (\NT - 1 ) $ matrix $ \uSigma $ is the matrix with the singular values of $ \uHb $.
Then, it can be shown that $ \uQ_2 = \mbI_{\NR} - \uHb \left( \uHbH \uHb \right)^{-1} \uHbH $ has the eigendecomposition $ \uQ_2 = \uUH \uLambda_{N} \uU $, where the diagonal $ \NR \times \NR $ matrix $ \uLambda_{N} $ has the $ N $ unit-valued eigenvalues of $ \uQ_2 $ grouped at the top-left on its main diagonal.
Since only $ \uU $ is random, the conditioning of $ \gamma_1 $ on $ \uQ_2 $ reduces to  the conditioning of $ \gamma_1 $ on $ \uU $.
Let us denote the normalized version of the $ \NR \times 1 $ deterministic vector $ \umu $ as $ \umu_{\text{n}} = \umu/\| \umu \| $.

Then,~(\ref{equation_gamma1_mgf_Turin}) yields
\begin{eqnarray}
\label{equation_gamma1_mgf_Turin_simplified_sdsger}
M_{\gamma_1|\uU}(s) = \frac{\exp \big\{ a \frac{\Gamma_1 s}{1 - \Gamma_1 s} \umu_{\text{n}}^{\cal{H}} \uU  \uLambda_{N} { \uUH \umu_{\text{n}} } \big \}}{\left( 1 - \Gamma_1 s \right)^N},
\end{eqnarray}
or
\begin{eqnarray}
\label{equation_gamma1_mgf_Turin_simplified}
M_{\gamma_1|\unu_{\text{n}}}(s) = \frac{1}{\left( 1 - \Gamma_1 s \right)^N} \exp \bigg\{ a \frac{\Gamma_1 s}{1 - \Gamma_1 s} \unu_{\text{n}}^{\cal{H}} \uLambda_{N} \unu_{\text{n}} \bigg \},
\end{eqnarray}
where
\begin{eqnarray}
\label{equation_a_first}
a = \RTinvaa \| \umu \|^2, \quad  \unu_{\text{n}} = \uUH \umu_{\text{n}}.
\end{eqnarray}
Note that the $ \NR \times 1 $ vector $ \unu_{\text{n}} $ has unit norm.
Next, we discuss the distributions of vector $ \unu_{\text{n}} $ and Hermitian form $ \unu_{\text{n}}^{\cal{H}} \uLambda_{N} \unu_{\text{n}} $ when $ \uHb $ is random.

\subsection{Special Case: Rician--Rayleigh Fading}
\label{section_special_case}

The above derivations are for the general case when any element of the channel matrix may experience Rician fading.
Let us now consider the special case of Rician--Rayleigh fading, whereby intended Stream 1 may experience Rician fading whereas interfering streams $ k = 2 : \NT $ experience Rayleigh fading, i.e., $ \uhda \neq \mzero $ and $ \uHdb = \mzero $ in~(\ref{equation_partitioned_H}).
Although this assumption reduces the generality of our results, it is required for tractability.
It also appears in Bartlett's decomposition theorem discussed in Section~\ref{section_Bartlett_only}.

Since matrix $ \uHb $ is zero-mean complex-valued Gaussian distributed, matrix $ \uU $ is isotropically, or Haar distributed on the group of $ \NR \times \NR $ unitary matrices\cite[Appendix~A.2]{marzetta_tit_99}\cite[\S3]{james_ams_64}\cite[Appendix~A]{jiang_tit_11}.
Because $ \umu_{\text{n}} $ is deterministic and belongs to the set $ \Omega_{\NR} $  of unit-norm vectors, $ \unu_{\text{n}} = \uUH \umu_{\text{n}} $ is isotropically distributed on $ \Omega_{\NR} $\cite[Appendix~A.2]{marzetta_tit_99}.

It is known from\cite[Appendix~A.1]{marzetta_tit_99} that if $ \uz \sim {\cal{CN}} \left(\mzero, \mbI_{\NR} \right) $ then $ \frac{\uz}{\| \uz \| } $ is also isotropically distributed on $ \Omega_{\NR} $, i.e., $ \unu_{\text{n}} \,{\buildrel d \over =}\, \frac{\uz}{\| \uz \| } $.
Thus,
\begin{eqnarray}
\label{equation_Hermitian_form_ratio_1}
\unu_{\text{n}}^{\cal{H}} \uLambda_{N} \unu_{\text{n}} { \,{\buildrel d \over =}\, } \frac{\uz^{\cal{H}}}{\| \uz \|} \uLambda_{N} \frac{\uz}{\| \uz \|} = \frac{\uz^{\cal{H}} \uLambda_{N} \uz}{\| \uz \|^2} = \frac{\uz^{\cal{H}} \uLambda_{N} \uz}{\uz^{\cal{H}} \mbI_{\NR} \uz} = \eta.
\end{eqnarray}
Substituting~(\ref{equation_Hermitian_form_ratio_1}) in~(\ref{equation_gamma1_mgf_Turin_simplified}) yields
\begin{eqnarray}
\label{equation_gamma1_mgf_Turin_simplified_1}
 M_{\gamma_1|\unu_{\text{n}}}(s) & \, {\buildrel d \over =} \, & M_{\gamma_1|\eta}(s) \nonumber \\ & \, {\buildrel d \over =} \, & \frac{1}{\left( 1 - \Gamma_1 s \right)^N} \exp \bigg\{ a \frac{\Gamma_1 s}{1 - \Gamma_1 s} \eta \bigg \}.
\end{eqnarray}

\subsection{Averaging the M.G.F.~of the Conditioned SNR}
\label{section_Average_Conditional_mgf_SNR}
Averaging~(\ref{equation_gamma1_mgf_Turin_simplified_1}) over $ \eta $ yields
\begin{eqnarray}
\label{equation_gamma1_mgf_Turin_simplified_2}
M_{\gamma_1}(s) & = & \mathbb{E}_{\eta} \{ M_{\gamma_1|\eta}(s) \} \nonumber \\ & = & \frac{1}{\left( 1 - \Gamma_1 s \right)^N} M_{\eta}\left( a \frac{\Gamma_1 s}{1 - \Gamma_1 s} \right),
\end{eqnarray}
where $ M_{\eta}(t) $ is the m.g.f.~of $ \eta $, which is derived next.
Let us rewrite $ \eta $ from~(\ref{equation_Hermitian_form_ratio_1}) as follows:
\begin{eqnarray}
\label{equation_Hermitian_form_ratio}
\eta & = & \frac{\sum_{i = 1}^{N} |z_i|^2}{\sum_{i = 1}^{\NR} |z_i |^2} = \frac{\sum_{i = 1}^{N} |z_i|^2}{\sum_{i = 1}^{N} |z_i |^2 + \sum_{i = N+1}^{\NR} |z_i |^2 } \nonumber \\ & = & \frac{\frac{2 N}{2 (\NR - N)} \left[ \frac{\sum_{i = 1}^{N} |z_i |^2}{2N} \right]/ \left[\frac{\sum_{i = N+1}^{\NR} |z_i |^2}{2 (\NR - N)} \right]}{\frac{2 N}{2 (\NR - N)} \left[ \frac{\sum_{i = 1}^{N} |z_i |^2}{2N} \right]/ \left[\frac{\sum_{i = N+1}^{\NR} |z_i |^2}{2 (\NR - N)} \right] + 1}.
\end{eqnarray}
Note that $ \sum_{i = 1}^{N} |z_i |^2 \sim \chi^2_{2 N} $ and $ \sum_{i = N+1}^{\NR} |z_i |^2 \sim \chi^2_{2 (\NR - N)} $\cite[Ch.~18]{johnson_book_95}.
Because they are also independent, we have that\cite[Section~6.4.3, \S2]{kokoska_crc_book_00}
\begin{eqnarray}
\label{equation_F_distribution}
\left[ \frac{\sum_{i = 1}^{N} |z_i |^2}{2N} \right]/\left[\frac{\sum_{i = N+1}^{\NR} |z_i |^2}{2 (\NR - N)} \right] \sim {\cal{F}} (2 N, 2 (\NR - N)).
\end{eqnarray}
Therefore, the distribution of $ \eta $ from~(\ref{equation_Hermitian_form_ratio}) is\cite[Vol. 2, p.~327]{johnson_book_95}
\begin{eqnarray}
\label{equation_Hermitian_form_ratio_Beta}
\eta \sim Beta(N, \NR - N),
\end{eqnarray}
i.e., the m.g.f.~of $ \eta $ is\cite[Section~6.2.1]{kokoska_crc_book_00}
\begin{eqnarray}
\label{equation_Hermitian_form_ratio_Beta_mgf}
M_{\eta}(\sigma) = \sum_{n = 0}^{\infty} \underbrace{ \frac{\left( N \right)_n}{\left( \NR \right)_n} \frac{\sigma^n}{n!} }_{=A_n(\sigma)} = {}_1 \! F_1(N; \NR; \sigma), \quad \forall \sigma \in \mathbb{R},
\end{eqnarray}
which becomes $ 1 $ for $ \sigma = 0 $, and $ e^\sigma $ for $ N = \NR $.

Finally, replacing~(\ref{equation_Hermitian_form_ratio_Beta_mgf}) into~(\ref{equation_gamma1_mgf_Turin_simplified_2}) yields the m.g.f.~expression for the SNR of ZF in Rician--Rayleigh fading
\begin{eqnarray}
\label{equation_gamma1_mgf_final}
M_{\gamma_1}(s) = \frac{1}{\left( 1 - \Gamma_1 s \right)^N} \,_1F_1 \left(N; \NR; a \frac{\Gamma_1 s}{1 - \Gamma_1 s} \right).
\end{eqnarray}

\subsection{SNR M.G.F.~Dependence on Channel Mean and Transmit-Correlation}
\label{section_channel_mean_correlation_effects}

Note that $ M_{\gamma_1}(s) $ depends on $ \uRT $ and $ \uhda $ through $ \Gamma_1 = \Gamma_{\text{s}}/\RTinvaa $ and $ a = \RTinvaa \| \umu \|^2 $, where $ \RTinvaa = (K + 1) \RTaloneinvaa $.
Using in~(\ref{equation_umu}) the assumption that the interfering streams experience Rayleigh fading, i.e., $ \uHdb = \mzero $, yields $ \umu = \uhad $. Thus, we have
\begin{eqnarray}
\label{equation_mu_norm}
\| \umu \|^2 & = & \| \uhad \|^2 = \| ( \uhad \; \mzero_{\NR \times (\NT - 1)} ) \|^2 = \| \uHd \|^2 \nonumber \\ & = & \frac{K}{K + 1} \NR \NT.
\end{eqnarray}
Based on~(\ref{equation_mu_norm}), we can write
\begin{eqnarray}
\label{equation_a_again}
a & = & K \NR \NT \RTaloneinvaa \propto K \NR \NT, \\
\label{equation_direction}
\uhad & = & \ud_{\text{n}} \sqrt{\frac{K}{K + 1} \NR \NT},
\end{eqnarray}
where $ \ud_{\text{n}} $ is the unit-norm vector that characterizes the `direction' of $ \uhda $.
Thus, the ZF SNR m.g.f.~in~(\ref{equation_gamma1_mgf_final}) is affected by vector $ \uhad $ only through its norm $ \| \uhad \| $, i.e., not through its `direction' vector $  \ud_{\text{n}}  $, and by the transmit-correlation matrix only through scalar $ \RTaloneinvaa $.



\subsection{ZF SNR M.G.F.~in Closed-Form}
\label{section_mgf_MIMO_ZF_SNR_Daalhuis}

We note that a closed-form expression for $ \,_1F_1 \left(N; \NR; \sigma_1 \right) $ exists\footnote{As pointed to us by Prof. A. B. Olde Daalhuis, the author of\cite[Ch.~13]{NIST_book_10}.}, and can be written based on\cite[Eqs.~(13.2.8), p.~322, (13.2.42), p.~325]{NIST_book_10} as
\begin{eqnarray}
\label{equation_1F1_NIST_expression}
\,_1F_1 \left(N; \NR; \sigma \right) = \quad \quad\quad\quad\quad\quad\quad \quad\quad\quad \quad\quad\quad \quad\quad\nonumber \\
\;\frac{(-1)^{N} (\NR - 1)!  }{(\NR - N - 1)! } \sum_{k = 0}^{\NR - N - 1} {{\NR - N - 1} \choose k} (N)_k \, \sigma^{-N - k} \nonumber \\
\; + \frac{ (\NR - 1)!  }{(N - 1)! }  e^{\sigma} \!  \!  \sum_{k = 0}^{N - 1} \! \!  \!  {N - 1 \choose k} (\NR - N)_k (-\sigma)^{-\NR + N - k},
\end{eqnarray}
which requires $ \sigma \neq 0 $, as well as $ \NR - N - 1 \ge 0 $, i.e., $ \NT \ge 2 $.
Nevertheless, for $ \NT = 1 $, i.e., $ N = \NR - \NT + 1 = \NR $, we already know that $ \,_1F_1 \left(N; \NR; \sigma \right) = \,_1F_1 \left(\NR; \NR; \sigma \right) = e^\sigma $.
When substituted in~(\ref{equation_gamma1_mgf_final}),~(\ref{equation_1F1_NIST_expression}) yields the SNR m.g.f.~in closed-form as follows:
\begin{eqnarray}
\label{equation_gamma1_mgf_final_new}
M_{\gamma_1}(s) =
\begin{cases}
\frac{1}{\left( 1 - \Gamma_1 s \right)^N}  e^{\frac{a\Gamma_1 s}{1 - \Gamma_1 s} }, & \!\!\!\!\!\!\!\!\!\! \!\!\!\!\!\!\!\!\!\!\!\!\!\!\!\!  \NT = 1, \\
\frac{1}{\left( 1 - \Gamma_1 s \right)^N} \left[\frac{ (-1)^{N} (\NR - 1)!  }{(\NR - N - 1)! } \sum_{k = 0}^{\NR - N - 1}  {{\NR - N - 1} \choose k} \right. \\ \left.  \; \times (N)_k \left(\frac{a\Gamma_1 s}{1 - \Gamma_1 s} \right)^{-N - k} \right.
\\
\left. \; + e^{\frac{a\Gamma_1 s}{1 - \Gamma_1 s} }  \frac{ (\NR - 1)!  }{(N - 1)! } \sum_{k = 0}^{N - 1} {N - 1 \choose k} \right. \\ \left. \; \times (\NR - N)_k \left(-\frac{a\Gamma_1 s}{1 - \Gamma_1 s} \right)^{-\NR + N - k} \right], & \!\!\!\!\!\!\!\!\!\! \!\!\!\!\!\!\!\!\!\!\!\!\!\!\!\! \NT \ge 2.
\end{cases}
\end{eqnarray}
From it, we shall express the ZF AEP in Section~\ref{section_AEP_derivations}.
Unfortunately, from it we cannot express the probability density function (p.d.f.) of the SNR $ p_{\gamma_1} (t) $ in terms of finite series.
Therefore, next, we recast the SNR m.g.f.~expression~(\ref{equation_gamma1_mgf_final}) as an infinite-series expression that easily yields its p.d.f.

\subsection{ZF SNR Distribution is Infinite Linear Combination of  Gamma Distributions}
\label{section_infinite_lin_comb_Gamma}

\label{section_mgf_MIMO_ZF_SNR_again}
Based on~(\ref{equation_Hermitian_form_ratio_Beta_mgf}), we can write the hypergeometric-function term from~(\ref{equation_gamma1_mgf_final}) as
\begin{eqnarray}
\label{equation_gamma1_mgf_final_again_temp}
& & \,_1F_1 \left( N; \NR; a \frac{\Gamma_1 s}{1 - \Gamma_1 s} \right)
= \sum_{n = 0}^{\infty} \overbrace{ \frac{\left( N \right)_n}{\left( \NR \right)_n} \frac{a^n}{n!}}^{=A_n(a)} \left(  \frac{ s \Gamma_1 }{1 - s \Gamma_1} \right)^n
\nonumber \\ &&
= \sum_{n = 0}^{\infty} A_n(a) \left( -1 + \frac{ 1 }{1 - s \Gamma_1} \right)^n
\nonumber \\ &&
= \sum_{n = 0}^{\infty} A_n(a) \sum_{m = 0}^{n} {n \choose m} (-1)^m \left( \frac{1}{1 - s \Gamma_1} \right)^{n - m}, \nonumber
\end{eqnarray}
so that the ZF SNR m.g.f.~from~(\ref{equation_gamma1_mgf_final}) can be recast as
\begin{eqnarray}
\label{equation_gamma1_mgf_final_again}
M_{\gamma_1}(s) =  \sum_{n = 0}^{\infty} A_n(a) \sum_{m = 0}^{n} {n \choose m}   (-1)^{m} \underbrace{\frac{ 1 }{(1 - s \Gamma_1)^{N + n - m}}}_{=M_{n,m}(s)}.
\end{eqnarray}
Note that $ M_{n,m}(s) $ is the m.g.f.~of $ Gamma (N + n - m, \Gamma_1) $, whose p.d.f.~is\cite[Section~IV.D]{siriteanu_tvt_11}:
\begin{eqnarray}
\label{equation_gamma1_pdf_Gamma}
p_{m,n} (t)  = \frac{t^{(N + n - m) - 1} e^{-t/\Gamma_1}}{[(N + n - m) - 1]! \,  \Gamma_1^{N + n - m}}, \quad t \ge 0.
\end{eqnarray}
Thus, the ZF SNR p.d.f.~corresponding to the m.g.f.~from~(\ref{equation_gamma1_mgf_final_again}) is expressed as the following infinite linear combination of p.d.f.s of gamma distributions:
\begin{eqnarray}
\label{equation_gamma1_pdf_final}
p_{\gamma_1} (t) = \sum_{n = 0}^{\infty} A_n(a) \sum_{m = 0}^{n} {n \choose m}  (-1)^{m}
p_{m,n} (t), \quad t \ge 0.
\end{eqnarray}

When all streams $ k = 1 : \NT $ undergo Rayleigh fading, i.e., for $ K = 0 $ (so that $ \uRTK = \uRT $, and $ \Gamma_k = { \Gamma_{\text{s}} }/{{[\uRTinv]_{k,k}}  } $), we have from~(\ref{equation_a_again}) that $ a = 0 $, and then only the terms for $ n = m = 0 $ remain from~(\ref{equation_gamma1_mgf_final_again}) and~(\ref{equation_gamma1_pdf_final}), and yield the following, known m.g.f.~and p.d.f.~expressions for the ZF SNR on Stream $ k $\cite{gore_cl_02}\cite{kiessling_spawc_03}:
\begin{eqnarray}
\label{equation mgf_sum_Ray}
M_{\gamma_k}(s) & = & \frac{ 1 }{(1 - s \Gamma_1)^{N}}, \\
\label{equation_gamma1_pdf_Gamma_Ray}
p_{\gamma_k} (t)  & = & \frac{t^{N-1} e^{-t/\Gamma_1}}{(N - 1)! \,  \Gamma_1^{N}}, \quad t \ge 0.
\end{eqnarray}
Thus, $ \gamma_k \sim Gamma (N, \Gamma_k) $ for Rayleigh--Rayleigh fading\cite{gore_cl_02}\cite{kiessling_spawc_03}.
Note that~(\ref{equation mgf_sum_Ray}) can also be deduced directly from~(\ref{equation_gamma1_mgf_final}), by substituting $ a = 0 $ and using the identity $ {}_1 \! F_1(N; \NR; 0) = 1 $.

Finally, for transmit-uncorrelated Rayleigh--Rayleigh fading, i.e., $ \uRT = \mbI_{\NT} $, the SNR m.g.f.~for any stream $ k $ is described, based on~(\ref{equation mgf_sum_Ray}), by
\begin{eqnarray}
\label{equation mgf_sum_Ray_uncorrelated}
M_{\gamma_k}(s) = \frac{ 1 }{(1 - s \Gamma_{\text{s}})^{N}}, \; \text{i.e., } \gamma_k \sim Gamma(N, \Gamma_{\text{s}}).
\end{eqnarray}

\subsection{ZF SNR Moments}
\label{section_moments_MIMO_ZF_SNR}

By using in~(\ref{equation_gamma1_mgf_final}) the following $ {}_1 \! F_1(\cdot; \cdot; \cdot) $ derivative property\cite[Eq.~(13.3.15), p.~325]{NIST_book_10}
\begin{eqnarray}
\label{equation_derivatives_1F1}
\frac{d^p}{d \sigma^p} \, {}_1 \! F_1 \left(N; \NR; \sigma \right)  = \frac{(N)_p}{(\NR)_p} \, {}_1 \! F_1 \left(N + p; \NR + p; \sigma \right),
\end{eqnarray}
we have obtained, through some manipulation, closed-form expressions for the first two derivatives of $ M_{\gamma_1}(s) $.
From them, we have expressed in Table~\ref{table_gamma1_statistics} the corresponding SNR moments as well as the SNR variance $ \mathbb{V} \{ \gamma_1 \} = \mathbb{E} \{ \gamma_1^2 \} - \left( \mathbb{E} \{ \gamma_1 \} \right)^2 $ and the amount of fading $ \mathbb{A} \{ \gamma_1 \} = \mathbb{V} \{ \gamma_1 \}/\left( \mathbb{E} \{ \gamma_1 \} \right)^2 $\cite[p.~18]{simon_alouini_book_00}, for Rician--Rayleigh and Rayleigh--Rayleigh fading.

\begin{table}
\caption{Moments, variance, and amount of fading for $ \gamma_1 $}
\renewcommand{\arraystretch}{1.35}
\label{table_gamma1_statistics}
\centering
\begin{tabular}{|c||c|c|}
  \hline
    & Rician & Rayleigh \\
   \hline \hline
   $ \mathbb{E} \{ \gamma_1 \} $ & $ N \, \Gamma_1 \left( 1 + \frac{a}{\NR} \right) $ & $ N \, \Gamma_1 $ \\
  \hline
   $ \mathbb{E} \{ \gamma_1^2 \} $ & $ N (N + 1) \, \Gamma_1^2 \left[ \left(1 + \frac{a}{\NR} \right)^2 - \frac{a^2}{N^2_{\text{R}}} \frac{1}{ (\NR + 1) } \right] $ & $ N (N + 1) \, \Gamma_1^2 $ \\
  \hline
   $ \mathbb{V} \{ \gamma_1 \} $ & $ N \Gamma_1^2 \left[ \left( 1 + \frac{a}{\NR} \right)^2  - \frac{a^2}{N^2_{\text{R}}} \, \frac{N + 1} {(\NR + 1) } \right] $ & $ N \Gamma_1^2 $ \\
  \hline
   $ \mathbb{A} \{ \gamma_1 \} $ & $ \frac{1}{N} \left[ 1 - \frac{N + 1}{\NR + 1} \frac{a^2}{(a + \NR)^2} \right] $ & $ \frac{1}{N} $ \\
  \hline
\end{tabular}
\end{table}

Using~(\ref{equation_gamma1_mgf_final}) and~(\ref{equation_derivatives_1F1}) to derive closed-form expressions for SNR moments of order $ p = 3, 4, \ldots $ becomes increasingly tedious.
Using the closed-form m.g.f.~expression in~(\ref{equation_gamma1_mgf_final_new}) is hardly more helpful.
On the other hand, from our alternative SNR m.g.f.~expression in~(\ref{equation_gamma1_mgf_final_again}) we can easily express the derivative of any order $ p $ of $ M_{\gamma_1}(s) $ in series form as
\begin{eqnarray}
\label{equation_gamma1_mgf_final_pth_derivative}
M^{(p)}_{\gamma_1}(s) = \quad \quad \quad \quad \quad \quad \quad \quad \quad \quad \quad \quad \quad \quad \quad \quad \quad \nonumber \\ \Gamma_1^p \sum_{n = 0}^{\infty} A_n(a) \sum_{m = 0}^{n} {n \choose m}   (-1)^{m} \frac{(N + n - m)_p}{(1 - s \Gamma_1)^{N + n - m + p}},
\end{eqnarray}
which yields the following expression for the moment of order $ p $ of $ \gamma_1 $:
\begin{eqnarray}
\label{equation_gamma1_mgf_final_pth_moment}
\mathbb{E} \{ \gamma_1^p \} = M^{(p)}_{\gamma_1}(0) \quad \quad \quad \quad \quad \quad \quad \quad \quad \quad \quad \quad \quad \quad \quad \quad \nonumber \\ = \Gamma_1^p \sum_{n = 0}^{\infty} A_n(a) \sum_{m = 0}^{n} {n \choose m} (-1)^{m} (N + n - m)_p.
\end{eqnarray}


\subsection{ZF Array Gain, Diversity Order, and Diversity Gain}
\label{section_diversity_order}

The improvement in average-SNR is known as array gain\cite[Eq.~(27)]{siriteanu_tvt_09}, and is reflected by a left-shift of the plot AEP-vs.-$ \Gamma_{\text{b}} $ at large $  { \Gamma_{\text{b}} }  $.
The top row in Table~\ref{table_gamma1_statistics} along with~(\ref{equation_a_again}) reveal that Rician fading yields an array gain of $ 1 + \frac{a}{\NR} = 1 + K \NT \RTaloneinvaa $ vs.~Rayleigh fading.

On the other hand, a performance improvement due to a higher magnitude of the slope of, e.g., AEP-vs.-$ \Gamma_{\text{b}} $ at large $  { \Gamma_{\text{b}} }  $ is known as diversity gain\cite{siriteanu_tvt_09}.
This slope-magnitude is known as diversity order.
According to\cite[Prop.~3]{wang_tcom_03}, under mild conditions on the error-probability dependence on SNR\cite[Sec.~II]{wang_tcom_03}, a scheme whose SNR m.g.f.~satisfies
\begin{eqnarray}
\label{equation_mgf_array_gain_div_order}
\lim_{s \to \infty } |M_{\gamma_1}(s)| \propto \frac{1}{s^N} + {\cal{O}} \left(\frac{1}{s^{N + 1}}\right)
\end{eqnarray}
has diversity order $ N $.
Using an approach similar to that from\cite[Example~3]{wang_tcom_03}, it can be shown that the ZF SNR m.g.f.~expression from~(\ref{equation_gamma1_mgf_final_again}) leads to~(\ref{equation_mgf_array_gain_div_order}).
Thus, ZF has diversity order $ N = \NR - \NT + 1 $ when the intended stream experiences either Rician or Rayleigh fading.
In conclusion, Rician fading yields array gain but no diversity gain vs.~Rayleigh fading.

\section{MIMO ZF Performance Analysis for Rayleigh--Rician Fading}
\label{section_Bartlett}

\subsection{Bartlett's Decomposition Theorem\cite[Sec.~III]{jayaweera_wpmc_03}}
\label{section_Bartlett_only}

For analysis tractability, Section~\ref{section_special_case} assumed Rician--Rayleigh fading, i.e., $ \uHd = (\uhda \; \mzero) $.
For the same assumptions, Bartlett's decomposition theorem, given below, characterizes the distributions of elements of the triangular decomposition of the noncentral-Wishart-distributed matrix $ \uW = \uHH \uH $.
As shown subsequently, this theorem helps characterize the ZF SNR distribution for Rayleigh--Rician fading, under certain restrictions.
These restrictions are then relaxed by applying analysis results from Section~\ref{section_performance_analysis} for a special mean--correlation relationship.

\begin{theorem}[Bartlett's Decomposition{\cite[Sec.~III]{jayaweera_wpmc_03}}]
\label{theorem_Bartlett}
Consider an $ \NR \times \NT $, complex-valued, Gaussian-distributed matrix $ \uH $ with mean $ \uHd = (\uhda \; \mzero_{\NR \times (\NT-1)}) $, row-correlation matrix $ \mbI_{\NR} $, and column-correlation matrix $ \RT = \mbI_{\NT} $. Then, the $ \NT \times \NT $ matrix $ \uW = \uHH \uH $ can be decomposed as $ \uW = \uTH \uT $, where $ \uT $ is an $ \NT \times \NT $ upper-triangular matrix with positive diagonal elements. Further, the nonzero elements of random matrix $ \uT $ are distributed as follows:
\begin{itemize}
\item $ [\uT]_{k,l} $ are mutually independent, $ \forall k, l = 1 : \NT $ such that $ k \le l $,
\item $ [\uT]_{k,l} \sim {\cal{N}}_{\text{c}}(0, 1) $, $ \forall k, l = 1 : \NT $ such that $ k < l $,
\item $ | [\uT]_{1,1} |^2 \sim \chi_{2\NR}^2(\|\uhda \|^2) $,
\item $ | [\uT]_{k,k} |^2 \sim \chi_{2(\NR - k + 1)}^2 $, $ \forall k = 2 : \NT $.
\end{itemize}
\end{theorem}

\subsection{SNR Distribution for Rayleigh Stream when at most One Interfering Stream is Rician}

Because the transmit-correlation matrix for our channel matrix $ \uH $ defined in~(\ref{equation channelH}) has been written as $ \uRTK = \frac{1}{K+1} \RT $, the Bartlett decomposition of the noncentral-Wishart-distributed matrix $ \uW = \uHH \uH $ from the conventional ZF SNR expression~(\ref{equation_gammak_perfect_CSI}) can be written as $ \uW = \frac{1}{K+1} \uTH \uT $.
This yields $ \uWinv = (K+1)\uTinv \uTinvH $, with $ \uTinv $ upper-triangular and $ \uTinvH $ lower-triangular, which
can be used to show that $ [\uWinv]_{\NT, \NT} = (K+1)/|[\uT]_{\NT,\NT}|^2 $.
Thus, the ZF SNR for Stream $ \NT $, which experiences Rayleigh fading, is described, based on~(\ref{equation_gammak_perfect_CSI}) and the last statement in Theorem~\ref{theorem_Bartlett}, by:
\begin{eqnarray}
\label{equation_gamma_NT_from_Bartlett}
\frac{K + 1}{ { \Gamma_{\text{s}} } } \, \gamma_{\NT} & = & \! \frac{K+1}{[\uWinv]_{\NT, \NT}} \nonumber \\ & = & \! |[\uT]_{\NT,\NT}|^2 \sim \chi_{2 (\NR - \NT + 1)}^2 = \chi_{2 N}^2
\end{eqnarray}

For the other Rayleigh-fading streams, i.e., $ k = 2 : \NT - 1 $, the SNRs are described by
\begin{eqnarray}
\label{equation_gamma_i_from_Bartlett}
\frac{K + 1}{ { \Gamma_{\text{s}} } } \, \gamma_k = \frac{K+1}{[\uWinv]_{k, k}} \neq |[\uT]_{k,k}|^2.
\end{eqnarray}
Thus, for $ k = 2 : \NT - 1 $, the distributions of $ \gamma_k $ cannot be characterized based on the distributions of $ |[\uT]_{k,k}|^2 $
provided by the Bartlett decomposition theorem.
Nevertheless, by symmetry, all the Rayleigh-fading streams must have the same distribution type, so that from~(\ref{equation_gamma_NT_from_Bartlett}) we deduce that
\begin{eqnarray}
\label{equation_gamma_2_NT_from_Bartlett}
\frac{K + 1}{ { \Gamma_{\text{s}} } } \, \gamma_k = \frac{K+1}{[\uWinv]_{k, k}}  \sim \chi_{2N}^2, \; \forall k = 2 : \NT.
\end{eqnarray}

A property analogous to~(\ref{equation_gamma_NT_from_Bartlett}) also does not hold for the Rician-fading Stream 1, i.e.,
\begin{eqnarray}
\label{equation_gamma_1_from_Bartlett}
\frac{K + 1}{ { \Gamma_{\text{s}} } } \, \gamma_1 = \frac{K + 1}{[\uWinv]_{1, 1}} \neq |[\uT]_{1,1}|^2.
\end{eqnarray}
As a consequence, Bartlett's decomposition cannot help characterize the ZF SNR distribution for the Rician-fading stream when the remaining streams undergo Rayleigh fading.
On the other hand, our earlier analysis successfully characterized the distribution of the ZF SNR $ \gamma_1 $ for the Rician-fading stream  as an infinite linear combination of gamma distributions, in~(\ref{equation_gamma1_mgf_final_again}).

\begin{remark}
\label{remark_1}
Bartlett's decomposition has revealed through~(\ref{equation_gamma_2_NT_from_Bartlett}) that, if Rician fading affects only one stream, and the fading is uncorrelated among all streams, then the ZF SNR for any of the Rayleigh-fading streams (with index $ k $) is characterized by
\begin{eqnarray}
\label{equation_gamma_2_NT_from_Bartlett_1}
\gamma_k \sim Gamma \left(N, {\frac{ { \Gamma_{\text{s}} } }{K+1}} \right).
\end{eqnarray}
Comparing~(\ref{equation_gamma_2_NT_from_Bartlett_1}) with~(\ref{equation mgf_sum_Ray_uncorrelated}) reveals for $ \RT = \mbI_{\NT} $ that, if $ \NT - 1 $ streams experience Rayleigh fading, then their ZF SNR distribution does not change type (from gamma) when the fading on the remaining stream changes from Rayleigh to Rician, and it is independent of $ \uhda $.
\end{remark}


\subsection{SNR Distribution for Rayleigh Stream when All Interfering Streams may be Rician}
\label{section_subsection_my_result_Ray_Rice}

Since the exponential term in the SNR m.g.f.~expression~(\ref{equation_gamma1_mgf_Turin}) disappears for $ \umu = \mzero $, i.e., for
\begin{eqnarray}
\label{equation_hd1_Hd2_r_cond}
 \uhda = \uHdb \ur_{2,1},
\end{eqnarray}
the distribution of $ \gamma_1 $ under this mean--correlation relationship is described simply by
\begin{eqnarray}
\label{equation_gamma1_mgf_Turin_simplified_umu_zero}
\label{equation_gamma1_distribution_umu_zero}
& \! \! \! \! \! \! M_{\gamma_1}(s) \! = \! \frac{1}{\left( 1 - \Gamma_1 s \right)^N},
\gamma_1 \! \sim \! Gamma \left(N, \Gamma_1 = \frac{ { \Gamma_{\text{s}} } }{\RTinvaa} \right)
\end{eqnarray}

\begin{remark}
\label{remark_2}
Condition~(\ref{equation_hd1_Hd2_r_cond}) holds if the following conditions both hold:
\begin{itemize}
\item $ \uhda = \mzero $, i.e., Stream 1 experiences Rayleigh fading,
\item $ \ur_{2,1} = \mzero$, i.e., the fading on Stream 1 is uncorrelated with the fading on other streams.
\end{itemize}
Then, the SNR for the Rayleigh-fading Stream 1 has the simple gamma distribution from~(\ref{equation_gamma1_distribution_umu_zero}).
\end{remark}

Thus, our analysis has revealed that the ZF SNR is gamma-distributed for Rayleigh-fading streams that are not correlated with the fading on any other stream, even when those other streams may experience transmit-correlated Rician fading.
On the other hand, Bartlett's decomposition theorem has helped characterize as gamma-distributed the ZF SNR of Rayleigh fading streams only when a single other stream may be Rician-fading, and the fading on all streams is uncorrelated --- see Remark~\ref{remark_1}.

\section{Performance--Measure Expressions for ZF}
\label{section_Performance_Measure_Expressions}

\subsection{Exact ZF AEP Expressions}
\label{section_AEP_derivations}

\label{section_AEP_ZF_correlated_estimated_derivation}
When the SNR m.g.f.~expression is available, one can apply the elegant AEP-derivation procedure from\cite[Ch.~9]{simon_alouini_book_00}, e.g., for $ M $PSK modulation (the same procedure also applies for other modulations).
Given $ \gamma_1 $, the error probability for Stream 1 can be written as\cite[Eq.~(8.22)]{simon_alouini_book_00}
\begin{eqnarray}\label{equation_instantaneous_Pe}
P_{\text{e}}(\gamma_1) = \frac{1}{\pi} \int_{0}^{\frac{M-1}{M}\pi}
\exp\left \{{-\gamma_1 \, \frac{ \sin^2{\frac{\pi}{M}}}{\sin^2\theta}} \right \} \mathrm{d} \theta.
\end{eqnarray}
Then, the AEP can be written in terms of the m.g.f.~of $ \gamma_1 $ as follows\cite[Chapter~9]{simon_alouini_book_00}:
\begin{eqnarray}
\label{equation_average_Pe}
P_{\text{e},1} = \mathbb{E}_{\gamma_1 } \{ P_{\text{e}}(\gamma_1) \}  = \frac{1}{\pi} \int_{0}^{\frac{M-1}{M}\pi}
M_{\gamma_1}\left(-\frac{ \sin^2{\frac{\pi}{M}}}{\sin^2 \theta}\right) \mathrm{d} \theta.
\end{eqnarray}
Based on~(\ref{equation_average_Pe}) and the SNR m.g.f.~expressions derived earlier, we provide below three alternative expressions for the AEP of ZF under Rician--Rayleigh fading.

First, substituting m.g.f.~expression~(\ref{equation_gamma1_mgf_final}) into~(\ref{equation_average_Pe}) yields the following ZF AEP expression, in terms of the confluent hypergeometric function:
\begin{eqnarray}
\label{equation_aep}
& P_{\text{e},1}  = \frac{1}{\pi} \int_0^{\frac{M-1}{M} \pi} \left( \frac{ \sin^{2}\theta }{ \sin^{2}\theta + \Gamma_1 \sin^2\frac{\pi}{M}} \right)^{N} \nonumber \\ & \quad \quad \quad \quad \quad \quad  \times {}_1 \! F_1 \left(N; \NR; -  \frac{a\Gamma_1 \sin^2\frac{\pi}{M}}{\sin^{2}\theta + \Gamma_1 \sin^2\frac{\pi}{M}} \right)  \mathrm{d} \theta.
\end{eqnarray}
Second, substituting~(\ref{equation_gamma1_mgf_final_new}) in~(\ref{equation_average_Pe}) yields the following ZF AEP expression, as a finite-limit integral of basic functions:
\begin{eqnarray}
\label{equation_aep_closed_form}
\!\!\! P_{\text{e},1} =
\begin{cases}
\frac{1}{\pi} \int_0^{\frac{M-1}{M} \pi} \left( \frac{ \sin^{2}\theta }{ \sin^{2}\theta + \Gamma_1 \sin^2\frac{\pi}{M}} \right)^{N} \\  \; \times\exp
\left(- \frac{a\Gamma_1 \sin^2\frac{\pi}{M}}{\sin^{2}\theta + \Gamma_1 \sin^2\frac{\pi}{M}} \right) \mathrm{d} \theta, & \!\!\! \!\!\! \!\!\! \!\!\!  \NT = 1,
\\
\frac{1}{\pi} \int_0^{\frac{M-1}{M} \pi} \left( \frac{ \sin^{2}\theta }{ \sin^{2}\theta + \Gamma_1 \sin^2\frac{\pi}{M}} \right)^{N} \\ \; \times \left[ \frac{(-1)^N (\NR - 1)!  }{(\NR - N - 1)! } \sum_{k = 0}^{\NR - N - 1} {{\NR - N - 1} \choose k} (N)_k  \right. \\ \; \quad \left. \times \left( - \frac{a\Gamma_1 \sin^2\frac{\pi}{M}}{\sin^{2}\theta + \Gamma_1 \sin^2\frac{\pi}{M}} \right)^{-N - k}  \right.
\\ \; \quad + \left. \exp
\left(- \frac{a\Gamma_1 \sin^2\frac{\pi}{M}}{\sin^{2}\theta + \Gamma_1 \sin^2\frac{\pi}{M}} \right) \frac{ (\NR - 1)!  }{(N - 1)! } \right. \\
\left. \; \quad \times \sum_{k = 0}^{N - 1} {N - 1 \choose k} (\NR - N)_k \right. \\ \left. \; \quad \times \left( \frac{a\Gamma_1 \sin^2\frac{\pi}{M}}{\sin^{2}\theta + \Gamma_1 \sin^2\frac{\pi}{M}} \right)^{-\NR + N - k} \right] \mathrm{d} \theta, & \!\!\! \!\!\! \!\!\! \!\!\! \NT \ge 2.
\end{cases}
\end{eqnarray}
Third, substituting~(\ref{equation_gamma1_mgf_final_again}) into~(\ref{equation_average_Pe}) yields the following ZF AEP expression, in terms of an infinite sum of finite-limit integrals:
\begin{eqnarray}
\label{equation_aep_sum}
\!\!\! \!\!\! P_{\text{e},1} & = &
\sum_{n = 0}^{\infty} A_n(a) \sum_{m = 0}^{n} {n \choose m}  (-1)^{m} \nonumber \\ && \: \times \frac{1}{\pi} \int_0^{\frac{M-1}{M} \pi} \! \! \!
\left( \frac{ \sin^{2}\theta }{ \sin^{2}\theta + \Gamma_1 \sin^2\frac{\pi}{M}} \right)^{N +n - m} \! \! \! \! \! \! \mathrm{d} \theta.
\end{eqnarray}

Numerical testing of these three AEP expressions has revealed that~(\ref{equation_aep_closed_form}) computes fastest but is inaccurate for $ \NT \ge 2 $ when $ \Gamma_{\text{b}} $ is small, because $ \Gamma_1 \propto \Gamma_{\text{b}} $ appears at negative powers in~(\ref{equation_aep_closed_form}).
Nevertheless,~(\ref{equation_aep_closed_form}) could be employed to quickly measure diversity gain and array gain, at high $ \Gamma_{\text{b}} $.
Since \texttt{MATLAB} provides the function \texttt{hypergeom} for $ {}_1 \! F_1 (\cdot; \cdot; \cdot)$, we will show AEP results for Rician--Rayleigh fading obtained only with (\ref{equation_aep}).

On the other hand, for the Rayleigh--Rician fading case discussed in Section~\ref{section_subsection_my_result_Ray_Rice}, the SNR m.g.f.~expression~(\ref{equation_gamma1_mgf_Turin_simplified_umu_zero}) yields the following AEP expression:
\begin{eqnarray}
\label{equation_aep_sum_Rayleigh_Rician_Cond}
P_{\text{e},1} =
\frac{1}{\pi} \int_0^{\frac{M-1}{M} \pi}
\left( \frac{ \sin^{2}\theta }{ \sin^{2}\theta + \Gamma_1 \sin^2\frac{\pi}{M}} \right)^{N}
\mathrm{d} \theta.
\end{eqnarray}
Note that, in all the above AEP expressions, $ \Gamma_1 $ defined in~(\ref{equation_Gamma}) depends on $ K $ as follows:
\begin{eqnarray}
\label{equation_Gamma_detail}
\Gamma_1 = \frac{ \Gamma_{\text{s}} }{\RTinvaa }  = \frac{1}{K + 1} \frac{ \Gamma_{\text{s}} }{{[\uRTinv]_{1,1}}  }
\end{eqnarray}

Finally, for Rayleigh--Rayleigh fading, any of the above AEP expressions reduces to
\begin{eqnarray}
\label{equation_aep_sum_Ray}
P_{\text{e},1} =
\frac{1}{\pi} \int_0^{\frac{M-1}{M} \pi}
\left( \frac{ \sin^{2}\theta }{ \sin^{2}\theta + \Gamma_1 \sin^2\frac{\pi}{M}} \right)^{N}
\mathrm{d} \theta,
\end{eqnarray}
where from~(\ref{equation_Gamma_detail}) with $ K= 0 $ we obtain
\begin{eqnarray}
\label{equation_Gamma_detail_Ray}
\Gamma_1 = \frac{ \Gamma_{\text{s}} }{{[\uRTinv]_{1,1}}  }.
\end{eqnarray}

\subsection{Approximate AEP Expression\cite{siriteanu_tvt_11}}
\label{section_approx_AEP}

As mentioned in the Introduction, in\cite{siriteanu_tvt_11} we attempted to analyze ZF performance, for fading that may be Rician--Rician, by approximating the noncentral-Wishart-distributed matrix $ \uW $ that enters the ZF SNR ratio-form from~(\ref{equation_gammak_perfect_CSI}) with a virtual central-Wishart-distributed matrix $ \uWhat $ of equal mean.
The column-correlation matrix $ \widehat{\mbR}_{\text{T},K}  $ of the virtual Rayleigh-fading channel matrix $ \uHhat $ that yields the approximating central-Wishart-distributed matrix $ \uWhat = \uHhatH \uHhat $ can be deduced from the condition $ \mathbb{E} \{ \uWhat \} = \mathbb{E} \{ \uW \} $, as\cite[Eq.~(11)]{siriteanu_tvt_11}
\begin{eqnarray}
\label{equation_virtual_correlation_matrix}
\widehat{\mbR}_{\text{T},K} = \mbR_{\text{T},K} + \frac{1}{\NR} \uHdH \uHd.
\end{eqnarray}

Then, for the virtual zero-mean channel matrix $ \uHhat $, the ZF SNR for Stream 1 is distributed as follows:
\begin{eqnarray}
\label{equation_Gamma_1_hat}
\widehat{\gamma}_1 \sim  Gamma (N, \widehat{\Gamma}_1 ), \quad \widehat{\Gamma}_1 = \frac{ { \Gamma_{\text{s}} } }{\left[ \widehat{\mbR}_{\text{T},K}^{-1} \right]_{1,1} }.
\end{eqnarray}
This virtual SNR distribution has yielded the following AEP expression for $M$PSK\cite[Eq.~(39)]{siriteanu_tvt_11}:
\begin{eqnarray}
\label{equation_aep_approx_Ray}
\widehat{P}_{\text{e},1} =
\frac{1}{\pi} \int_0^{\frac{M-1}{M} \pi}
\left( \frac{ \sin^{2}\theta }{ \sin^{2}\theta + \widehat{\Gamma}_1  \sin^2\frac{\pi}{M}} \right)^{N}
\mathrm{d} \theta,
\end{eqnarray}
which has been used to approximate the actual AEP for Rician--Rician fading in\cite{siriteanu_tvt_11}.

Note that for Rician--Rayleigh fading, i.e., when $ \uHd = \left( \uhad \; \mzero_{\NR \times (\NT - 1)} \right)  $,
\begin{eqnarray}
\label{equation_virtual_correlation_matrix_1}
\widehat{\mbR}_{\text{T},K} & = & \mbR_{\text{T},K} + \frac{1}{\NR} \uHdH \uHd \nonumber \\ & = & \mbR_{\text{T},K} + \frac{1}{\NR}  \left(
  \begin{array}{cc}
    \| \uhda \|^2 & \mzero \\
    \mzero & \mzero \\
  \end{array}
  \right),
\end{eqnarray}
i.e., the distribution of the virtual SNR $ \widehat{\gamma}_1 $ from~(\ref{equation_Gamma_1_hat}) depends on $ \uhda $ only through its norm, like the actual SNR m.g.f.~$ \gamma_1 $, as shown in Section~\ref{section_channel_mean_correlation_effects}.
Nevertheless, we shall see that this does not help the accuracy of the AEP approximation in~(\ref{equation_aep_approx_Ray}).

\subsection{Closed-Form of the Finite-Limit Integral in AEP Expressions\cite{siriteanu_tvt_09}\cite{simon_alouini_book_00}}

For the numerical results shown later, the finite-limit integral that appears in the AEP expressions~(\ref{equation_aep_sum_Rayleigh_Rician_Cond}),~(\ref{equation_aep_sum_Ray}), and~(\ref{equation_aep_approx_Ray}) has been computed in \texttt{MATLAB} by using the following closed-form expression obtained in\cite[Eq.~(33)]{siriteanu_tvt_09} from\cite[Eqs. (5A.17-19), pp. 127-128]{simon_alouini_book_00}:
\begin{eqnarray}
\label{equation_finite_limit_integral}
\frac{1}{\pi} \int_0^{\frac{M-1}{M} \pi}
\left( \frac{ \sin^{2}\theta }{ \sin^{2}\theta + \Gamma_1 \sin^2\frac{\pi}{M}} \right)^{N}
\mathrm{d} \theta \quad \quad \quad \quad \quad \nonumber \\ \; = \frac{M - 1}{M} - \frac{b}{\pi} \,  \, \sum_{n = 0}^{N -
1} {2n \choose n} \, \left( \frac{1 - b^2}{4} \right)^n \quad \quad \quad \quad  \nonumber \\ \; \; \times \! \left\{ \! \frac{\pi}{2} + \varphi + \sin \varphi \sum_{i = 1}^{n} \frac{
4^{(n - i)} \left[ \cos \varphi \right]^{2 \, (n - i) + 1} }{{2(n - i) \choose
n - i} \, \left[ 2(n-i) + 1 \right]} \! \right\} \! \!,
\end{eqnarray}
where $ b
\stackrel{\triangle}{=} \sqrt{\frac{\Gamma_1 \sin^2\frac{\pi}{M}}{\Gamma_1 \sin^2\frac{\pi}{M} + 1}} $, and $ \varphi
\stackrel{\triangle}{=} \tan^{-1} \! (b/\tan \frac{\pi}{M}) $.


\subsection{Exact ZF Outage Probability and Ergodic Capacity Expressions}
\label{section_OP_capacity_MIMO_ZF_SNR}

The outage probability at threshold SNR $ \gamma_{1,\text{th}} $ is\cite[p.~5]{simon_alouini_book_00}:
\begin{eqnarray}
\label{equation_Po_definition}
P_{\text{o}}(\gamma_{1,\text{th}} ) = Pr ( \gamma_1 \le  \gamma_{1,\text{th}}) = \int_{0}^{\gamma_{1,\text{th}}} p_{\gamma_1 } (t) \mathrm{d} t.
\end{eqnarray}
Note that the outage probability is also the cumulative distribution function of the SNR.
Integrating~(\ref{equation_gamma1_pdf_final}) yields the following expression for the outage probability:
\begin{eqnarray}
\label{equation_gamma1_cdf_final}
P_{\text{o}} = \sum_{n = 0}^{\infty} A_n(a) \sum_{m = 0}^{n} {n \choose m}  (-1)^{m} \frac{\gamma \left( N + n - m,  \gamma_{1,\text{th}}/\Gamma_1 \right)}{[(N + n - m) - 1]!}
\end{eqnarray}

On the other hand, given the ZF SNR $ \gamma_1 $, the instantaneous capacity, measured in bits-per-channel-use (bpcu), is:
\begin{eqnarray}
\label{equation_capacity_gamma1}
C(\gamma_1) = \log_2 (1 + \gamma_1) = \frac{1}{\ln 2} \ln (1 + \gamma_1).
\end{eqnarray}
The average, or ergodic, capacity can be expressed by using the $ \gamma_1 $ p.d.f.~expression~(\ref{equation_gamma1_pdf_final}) as:
\begin{eqnarray}
\label{equation_capacity_ergodic_inf_sum}
C & = & \mathbb{E}_{\gamma_1 } \{ C(\gamma_1) \} = \int_{0}^{\infty} \frac{1}{\ln 2} \ln (1 + t) p_{\gamma_1} (t) \mathrm{d} t\nonumber \\
& = & \frac{1}{\ln 2} \sum_{n = 0}^{\infty} A_n(a) B_n = \frac{1}{\ln 2} \sum_{n = 0}^{\infty} T_n,
\end{eqnarray}
with $ B_n $ given by the sum\footnote{The inner-sum indexing over $ Q = N - 1 : n + N - 1$, as opposed to $ m = 0 : n $, may also be used in~(\ref{equation_gamma1_mgf_final_again}),~(\ref{equation_gamma1_pdf_final}), (\ref{equation_gamma1_cdf_final}).}
\begin{eqnarray}
\label{equation_Bn}
B_n = \sum_{Q = N - 1}^{n + N - 1} {n \choose n + N - 1 - Q}  (-1)^{n + N - 1 - Q} C_{Q}(\Gamma_1),
\end{eqnarray}
where we have derived $ C_{Q}(\Gamma_1) $ as follows\footnote{Expression\cite[Eq.~(4.222.8), p.~530]{gradshteyn_book_07} for $ C_{Q}(\cdot) $ is incorrect.}:
\begin{eqnarray}
\label{equation_capacity_ergodic_inf_sum_term}
C_{Q}(\Gamma_1) & = & \frac{ 1 }{ Q! } \frac{ 1 }{ \Gamma_1^{Q+ 1} }  \int_{0}^{\infty} \ln (1 + t) t^{Q} e^{-t/\Gamma_1} \mathrm{d} t
\nonumber \\ & = &
\frac{ 1 }{ Q! }  \int_{0}^{\infty} \ln (1 + \Gamma_1 y) y^{Q} e^{-y} \mathrm{d} y
\nonumber \\ & = & e^{\frac{1}{\Gamma_1}} \text{E}_1 \left(\frac{1}{\Gamma_1} \right)
\nonumber \\ && +
\sum_{q_1 = 1}^{Q} \frac{1}{q_1!} \left[ \left( - \frac{1}{\Gamma_1} \right)^{q_1} e^{\frac{1}{\Gamma_1}} \text{E}_1 \left(\frac{1}{\Gamma_1} \right) \right.
\nonumber \\ && \left. \; + \sum_{q_2 = 0}^{q_1 - 1}\left( - \frac{1}{\Gamma_1} \right)^{q_2} (q_1 - q_2 - 1)! \right].
\end{eqnarray}

\subsection{Infinite-Series Convergence\cite[Sec.~III]{siriteanu_ausctw_14}}
\label{section_convergence}

The fact that the confluent hypergeometric function series expression~(\ref{equation_Hermitian_form_ratio_Beta_mgf}) that enters the ZF SNR m.g.f.~expression~(\ref{equation_gamma1_mgf_final}) converges for any $ \sigma $ has been known\cite[Eq.~13.2.2, p.~322]{NIST_book_10}.
This assertion is supported by its alternative closed-form expression in~(\ref{equation_1F1_NIST_expression}).
Nevertheless, we have provided in\cite[Sec.~III.B]{siriteanu_ausctw_14} a proof obtained using the ratio-test theorem\cite[Th.~1.4, pp.~6-7]{andrews_book_85}.
Thus, also the ZF SNR m.g.f.~expression~(\ref{equation_gamma1_mgf_final}) converges.

As a result of the convergence of~(\ref{equation_Hermitian_form_ratio_Beta_mgf}) and~(\ref{equation_gamma1_mgf_final}), the AEP expressions~(\ref{equation_aep}) and~(\ref{equation_aep_sum}) converge\cite[Sec.~III.D]{siriteanu_ausctw_14}.
Further, the SNR p.d.f.~expression~(\ref{equation_gamma1_pdf_final}) converges, as proved by successively upper-bounding $ p_{\gamma_1} (t) $ in\cite[Sec.~III.C]{siriteanu_ausctw_14}.
Finally, the ergodic-capacity expression~(\ref{equation_capacity_ergodic_inf_sum}) converges, as proved by sum-splitting and successive upper-bounding in\cite[Sec.~III.E]{siriteanu_ausctw_14}.

\subsection{Infinite-Series Computation\cite[Sections~IV, V]{siriteanu_ausctw_14}}
\label{section_computation}

The magnitude of $ \sigma $ determines the ability to compute (i.e., approximate numerically) the infinite series in~(\ref{equation_Hermitian_form_ratio_Beta_mgf}), for any known computation method\cite{muller_nm_01}\cite{siriteanu_ausctw_14}.
Thus, the magnitude of $ a $ determines the ability to compute~(\ref{equation_gamma1_mgf_final}) and the ensuing infinite series derived above.
Since, from~(\ref{equation_a_again}), $ a \propto K \NR \NT  $, more series terms need to be considered for accurate computation when $ K $, $ \NR $, and $ \NT $ increase.
On the other hand, for example, computation of $ B_n $ with~(\ref{equation_Bn}) encounters numerical instability at lower $ n $, as $ K $ increases.
We explain these issues in detail in\cite[Sections~IV.B, V]{siriteanu_ausctw_14}.

To compute the infinite series for the SNR p.d.f., outage probability, and ergodic capacity from~(\ref{equation_gamma1_pdf_final}),~(\ref{equation_gamma1_cdf_final}), and~(\ref{equation_capacity_ergodic_inf_sum}), respectively, we adapted the fast and reliable recursive approach from\cite[Method~1]{muller_nm_01}.
It computes new series terms (e.g., $ T_n $ for the ergodic capacity expression) until the relative change (i.e., $ | T_n/\sum_{i = 0}^{n} T_i | $) is smaller than a tolerance level. It stops at $ n = 150 $ if numerical convergence does not occur earlier\footnote{The MIMO study in\cite{kim_tvt_12} yielded infinite-series ensuing from  $ {}_1 \! F_1\left(\cdot; \cdot; \cdot \right) $ and faced numerical issues similarly. Others have been less systematic\cite{dharmawansa_globecom_08}.}.
This method is discussed further in\cite[Sections~IV.A, V]{siriteanu_ausctw_14}.

As a result of the numerical issues mentioned above, there exist values of $ K $, $ \NR $, and $ \NT $ beyond which the series cannot be computed reliably with this method.
The argument range that allows for accurate computation of the series may be extendable by increasing representation precision, e.g., in \texttt{MATLAB}, with\cite{advanpix_soft_13}.
Adapting the computation method may also help: small arguments can be tackled with simple, i.e., fast, methods; larger arguments require involved, i.e., slow, methods; even larger arguments can be tackled only through series approximations\cite{muller_nm_01}.

\section{Numerical Results}
\label{section_Numerical_Results}

\subsection{Settings}
\label{section_Setting}

Numerical results obtained in \texttt{MATLAB} are presented for $ \NR = 4 $, $ \NT = 1 : 4 $, ZF for Stream 1, QPSK ($ M = 4 $), and relevant values and ranges of $  { \Gamma_{\text{s}} } $ and $  { \Gamma_{\text{b}} } $.
Correlation matrix $ \uRT $ has been computed as in\cite{siriteanu_tvt_11}, for a uniform linear antenna array with interelement distance normalized to carrier-half-wavelength $ d_{\text{n}} = 1 $, and Laplacian power azimuth spectrum centered at $ \theta_{\text{c}} = 5^\circ $. We have set $ K $ and AS to the averages of their lognormal distributions for WINNER scenarios A1 (indoor), C2 (typical urban macrocell), and D1 (rural macrocell)\cite[Table~I]{siriteanu_tvt_11}, unless specified otherwise.
Finally, $ \uhda $ has been computed with~(\ref{equation_direction}), for `direction' vector set to $ \ud_{\text{n}} = (1 \; 1 \; 1 \; 1)^{\cal{T}}/\sqrt{4} $, unless specified otherwise.

\subsection{Computation of AEP}
\label{section_approach_AEP}

The AEP has been computed from:
\begin{itemize}
\item New exact expression~(\ref{equation_aep}), using the \texttt{hypergeom} function\footnote{The code for this function is not accessible, i.e., there is no control over the number of terms used in computing $ F_1 \left(\cdot; \cdot; \cdot \right) $ for~(\ref{equation_aep}). Nevertheless, AEP computation has been accurate for all tried $ \NR $, $ \NT $, and $ K $ values.}, for Rician--Rayleigh fading:
\begin{itemize}
\item denoted in figures as `Rice--Ray' and `exact'.
\end{itemize}
\item New exact expression~(\ref{equation_aep_sum_Rayleigh_Rician_Cond}), for Rayleigh--Rician fading:
\begin{itemize}
\item denoted in figures as `Ray--Rice' and `exact'.
\end{itemize}
\item Known exact expression~(\ref{equation_aep_sum_Ray}), for Rayleigh--Rayleigh fading:
\begin{itemize}
\item denoted in figures as `Ray--Ray' and `exact'.
\end{itemize}
\item Known approximate expression~(\ref{equation_aep_approx_Ray}), for Rician--Rayleigh, Rayleigh--Rician, and Rayleigh--Rayleigh fading:
\begin{itemize}
\item denoted in figures as `Rice--Ray' or `Ray--Rice' or `Ray--Ray', and `approx'.
\end{itemize}
\item Monte Carlo simulation, using at least $ 10^6 $ samples, for all fading cases:
\begin{itemize}
\item denoted in figures as `Rice--Ray' or `Ray--Rice' or `Ray--Ray', and `sim'.
\end{itemize}
\end{itemize}
Note that the finite-limit integral that appears in~(\ref{equation_aep_sum_Rayleigh_Rician_Cond}),~(\ref{equation_aep_sum_Ray}), and~(\ref{equation_aep_approx_Ray}) has been computed with~(\ref{equation_finite_limit_integral}).

\subsection{Computation of $ p_{\gamma_1 } $, $ P_{\text{o}} $, and $ C $}
\label{section_approach_cap}

The series for the ZF SNR p.d.f., outage probability, and ergodic capacity from~(\ref{equation_gamma1_pdf_final}),~(\ref{equation_gamma1_cdf_final}), and~(\ref{equation_capacity_ergodic_inf_sum}) have been computed with the approach outlined in Section~\ref{section_computation}, with tolerance level $ 10^{-10} $ for $ p_{\gamma_1 } $ and $ P_{\text{o}} $, and $ 10^{-5} $ for $ C $.
For $ \NR = \NT = 4 $, we have found that the computation yields good agreement with the simulation only for $ K \le 1.2 $~dB, which allows the series to converge numerically for a smaller number of terms than the number that produces numerical instability.
(The attained number of terms, averaged over $ { \Gamma_{\text{b}} } $, is shown in the title of relevant figures as $ n_{\text{max}} $.)
We have outlined in Section~\ref{section_computation} and detailed in\cite[Sec.~V]{siriteanu_ausctw_14} explanations and possible solutions for these limitations.
Finally, function \texttt{expint} has been employed to compute $ \text{E}_1 \left( \cdot \right) $ for~(\ref{equation_capacity_ergodic_inf_sum_term}).\label{pageref_pdfsettings}

\subsection{Description of Results for Rician--Rayleigh and Rayleigh--Rayleigh Fading}

\begin{figure}[t]
\begin{center}
\includegraphics[width=3.4in]
{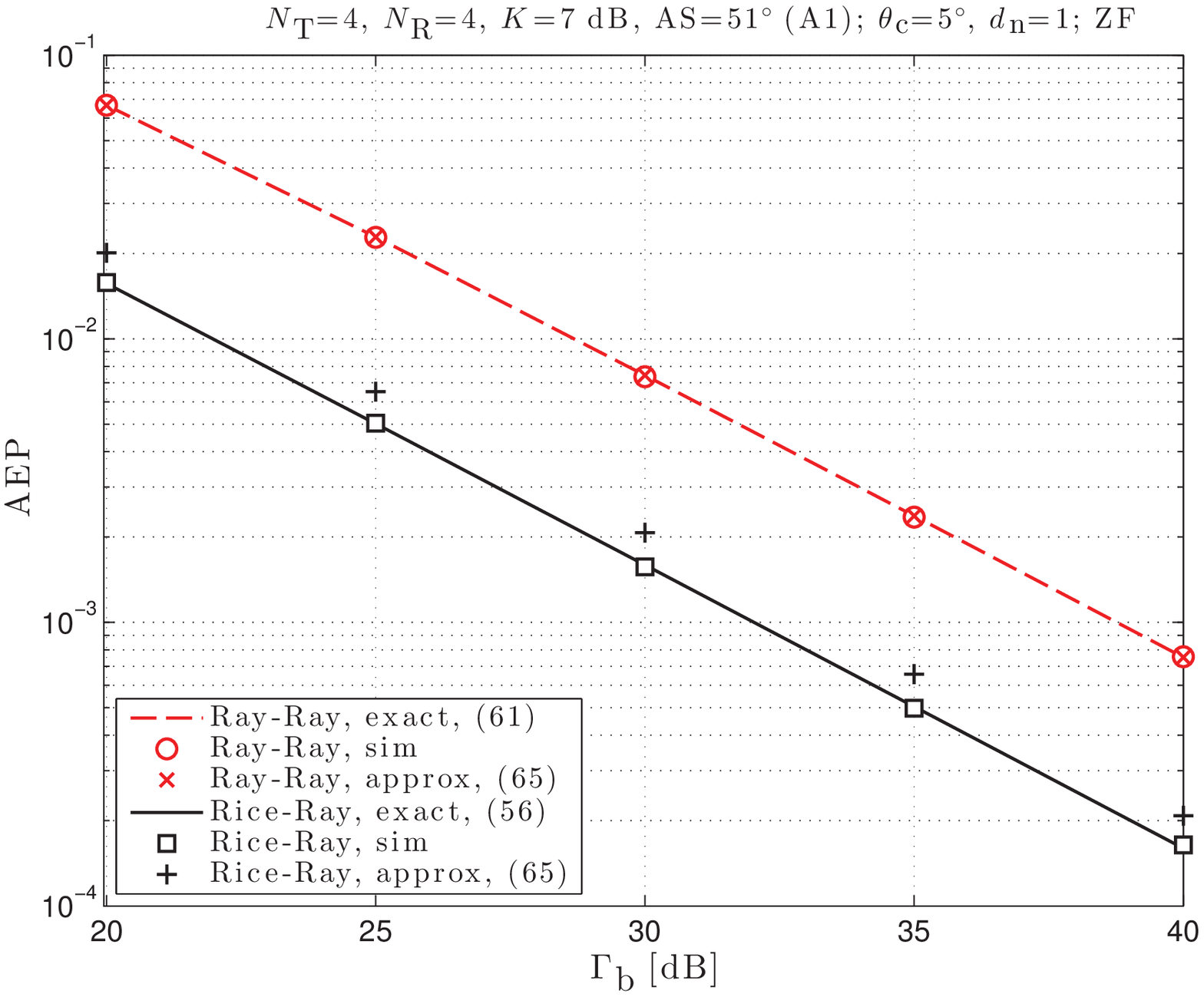}
\caption{Stream-1 AEP for $ \NR = \NT =4 $, $ K = 7 $~dB, $ \text{AS} = 51^{\circ} $ and $ \ud_{\text{n}} = (1 \; 1 \; 1 \; 1)^{\cal{T}}/\sqrt{4}  $.}
\label{figure_AEP_AER_vs_SNR_ZF_Rice_Ray_Fixed_AS_K_A1_NT4_NR4}
\end{center}
\end{figure}

Fig.~\ref{figure_AEP_AER_vs_SNR_ZF_Rice_Ray_Fixed_AS_K_A1_NT4_NR4} shows, for $ \NT = 4 $ and AS and $ K $ set to WINNER averages for scenario A1, close agreement for Rician--Rayleigh fading between the AEP from the new exact expression~(\ref{equation_aep}) and from simulation.
On the other hand, the AEP from the approximate expression~(\ref{equation_aep_approx_Ray}) underestimates the actual AEP by more than $ 1 $~dB over the entire $  { \Gamma_{\text{b}} } $ range.
We have obtained similar results (unshown) for scenarios with other combinations of average $ K $ and AS values, as well as for other $ \theta_{\text{c}} $ values.
Thus, for Rician--Rayleigh fading, AEP approximation accuracy with expression~(\ref{equation_aep_approx_Ray}) is largely independent of $ K $, AS, and $ \theta_{\text{c}} $.
On the other hand, we have found that that the accuracy of~(\ref{equation_aep_approx_Ray}) degrades with increasing $ \NR - \NT $  and with decreasing $ \NR = \NT  $.
For example, for $ \NR = 4 $ and $ \NT = 2 $, expression~(\ref{equation_aep_approx_Ray}) underestimates the actual AEP by nearly $ 3.5 $~dB.
On the other hand, for $ \NR = \NT = 3 $, expression~(\ref{equation_aep_approx_Ray}) underestimates the actual AEP by about $ 1.7 $~dB.
For Rician--Rician fading, in\cite{siriteanu_tvt_11}, we also found that the accuracy of~(\ref{equation_aep_approx_Ray}) degrades with increasing $ \NR - \NT $  and with decreasing $ \NR = \NT  $.
On the other hand, approximation accuracy there was found dependent on $ K $, AS, and $ \theta_{\text{c}} $.

\begin{figure}[t]
\begin{center}
\includegraphics[width=3.4in]
{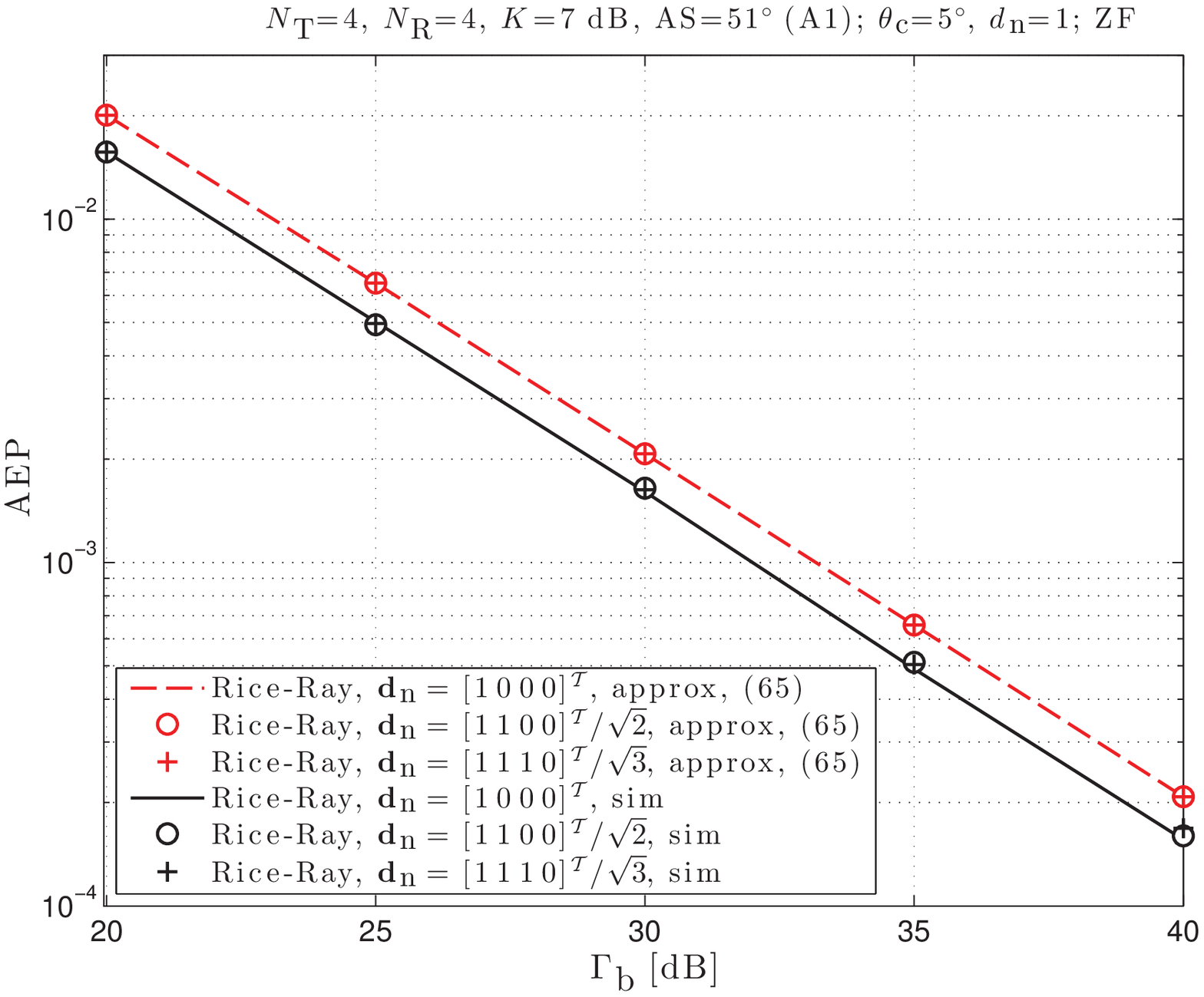}
\caption{Stream-1 AEP for $ \NR = \NT =4 $, $ K = 7 $~dB, $ \text{AS} = 51^{\circ} $, and various choices of $ \ud_{\text{n}} $.}
\label{figure_AEP_AER_vs_SNR_ZF_Rayleigh_Rice_Fixed_AS_K_A1_NT4_NR4_Hd}
\end{center}
\end{figure}

Fig.~\ref{figure_AEP_AER_vs_SNR_ZF_Rayleigh_Rice_Fixed_AS_K_A1_NT4_NR4_Hd} shows AEP results for Rician--Rayleigh fading, for $ \NT = 4 $ and various choices of the `direction' $ \ud_{\text{n}} $ of $ \uhda $.
The simulation results confirm that the actual ZF performance does not depend on this `direction'.
They agree with (unshown) results from the new exact expression~(\ref{equation_aep}).
They also agree with the results for $ \ud_{\text{n}} = (1 \; 1 \; 1 \; 1)^{\cal{T}}/\sqrt{4} $ from Fig.~\ref{figure_AEP_AER_vs_SNR_ZF_Rice_Ray_Fixed_AS_K_A1_NT4_NR4}.
Finally, Fig.~\ref{figure_AEP_AER_vs_SNR_ZF_Rayleigh_Rice_Fixed_AS_K_A1_NT4_NR4_Hd} confirms that the AEP approximation computed with~(\ref{equation_aep_approx_Ray}) is also independent of $ \ud_{\text{n}}  $.

\begin{figure}[t]
\begin{center}
\includegraphics[width=3.4in]
{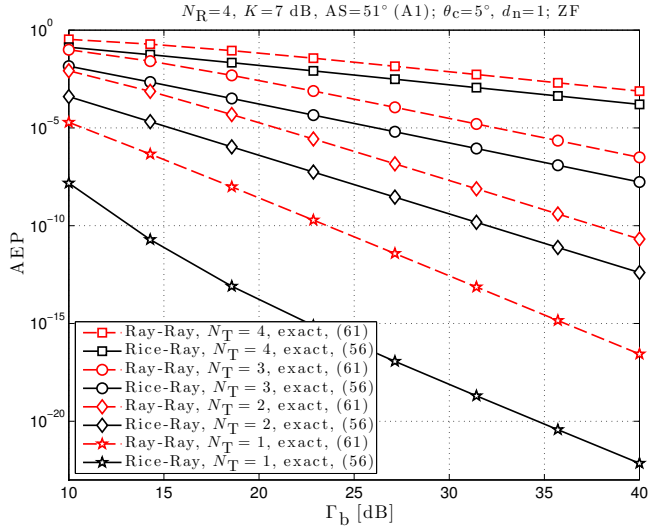}
\caption{Stream-1 AEP for $ \NR = 4 $, $ \NT = 1 : 4 $, $ K = 7 $~dB, $ \text{AS} = 51^{\circ} $.}
\label{figure_AEP_AER_vs_hSNR_ZF_Rayleigh_Rice_Fixed_AS_K_A1_NT_1_4_NR4}
\end{center}
\end{figure}

Figs.~\ref{figure_AEP_AER_vs_SNR_ZF_Rice_Ray_Fixed_AS_K_A1_NT4_NR4} and~\ref{figure_AEP_AER_vs_SNR_ZF_Rayleigh_Rice_Fixed_AS_K_A1_NT4_NR4_Hd} reveal that for $ \NR = \NT = 4 $ the ZF diversity order is $ N = \NR - \NT + 1 = 1 $ for both Rician--Rayleigh and Rayleigh--Rayleigh fading. Fig.~\ref{figure_AEP_AER_vs_hSNR_ZF_Rayleigh_Rice_Fixed_AS_K_A1_NT_1_4_NR4} confirms that the ZF diversity order is $ N $ for all $ \NT $ choices.
Finally, Figs.~\ref{figure_AEP_AER_vs_SNR_ZF_Rice_Ray_Fixed_AS_K_A1_NT4_NR4} and~\ref{figure_AEP_AER_vs_hSNR_ZF_Rayleigh_Rice_Fixed_AS_K_A1_NT_1_4_NR4} reveal that ZF for Stream 1 yields an array gain when this stream undergoes Rician fading instead of Rayleigh fading.

\begin{figure}[t]
\begin{center}
\includegraphics[width=3.4in]
{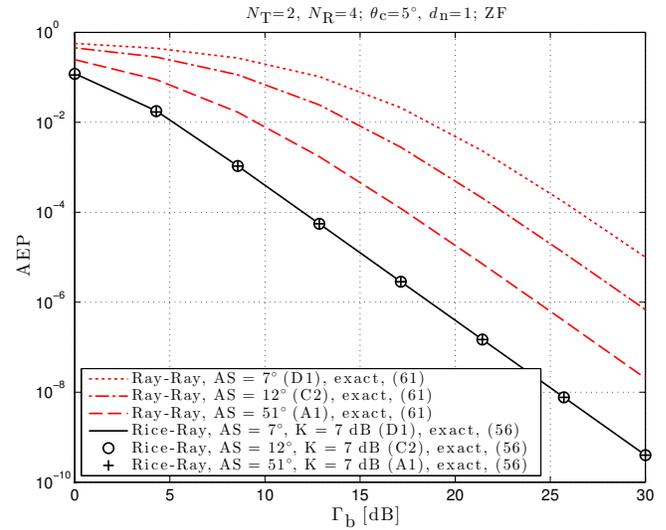}
\caption{Stream-1 AEP for $ \NR = 4 $, $ \NT = 2 $, and $ K $, AS set to averages for scenarios A1, C2, and D1.}
\label{figure_AEP_vs_SNR_ZF_Rice_Ray_A1_C2_D1_NT_2_NR_4}
\end{center}
\end{figure}

Fig.~\ref{figure_AEP_vs_SNR_ZF_Rice_Ray_A1_C2_D1_NT_2_NR_4} shows the AEP for $ \NT = 2 $, $ \NR = 4 $, and $ K $ and AS set to their averages for WINNER scenarios A1, C2, and D1.
Note that the average $ K $ for all these scenarios is $ 7 $~dB, which helps isolate the effect of AS variation.
When the intended stream experiences Rayleigh fading, the AEP decreases with increasing transmit AS, i.e., with decreasing transmit-correlation.
This performance improvement is due exclusively to array gain, since the diversity order is $ N = 3 $ for all cases.
Thus, when the intended stream experiences Rician fading with $ K $ near its WINNER average, the AEP is largely unaffected by AS.
Unshown simulation results have confirmed this finding.
We have found that the AEP is affected by AS only for $ K < 0 $~dB, which occurs relatively infrequently in practice, according to WINNER\cite[Table~I]{siriteanu_tvt_11}.

\begin{figure}[t]
\begin{center}
\includegraphics[width=3.4in]
{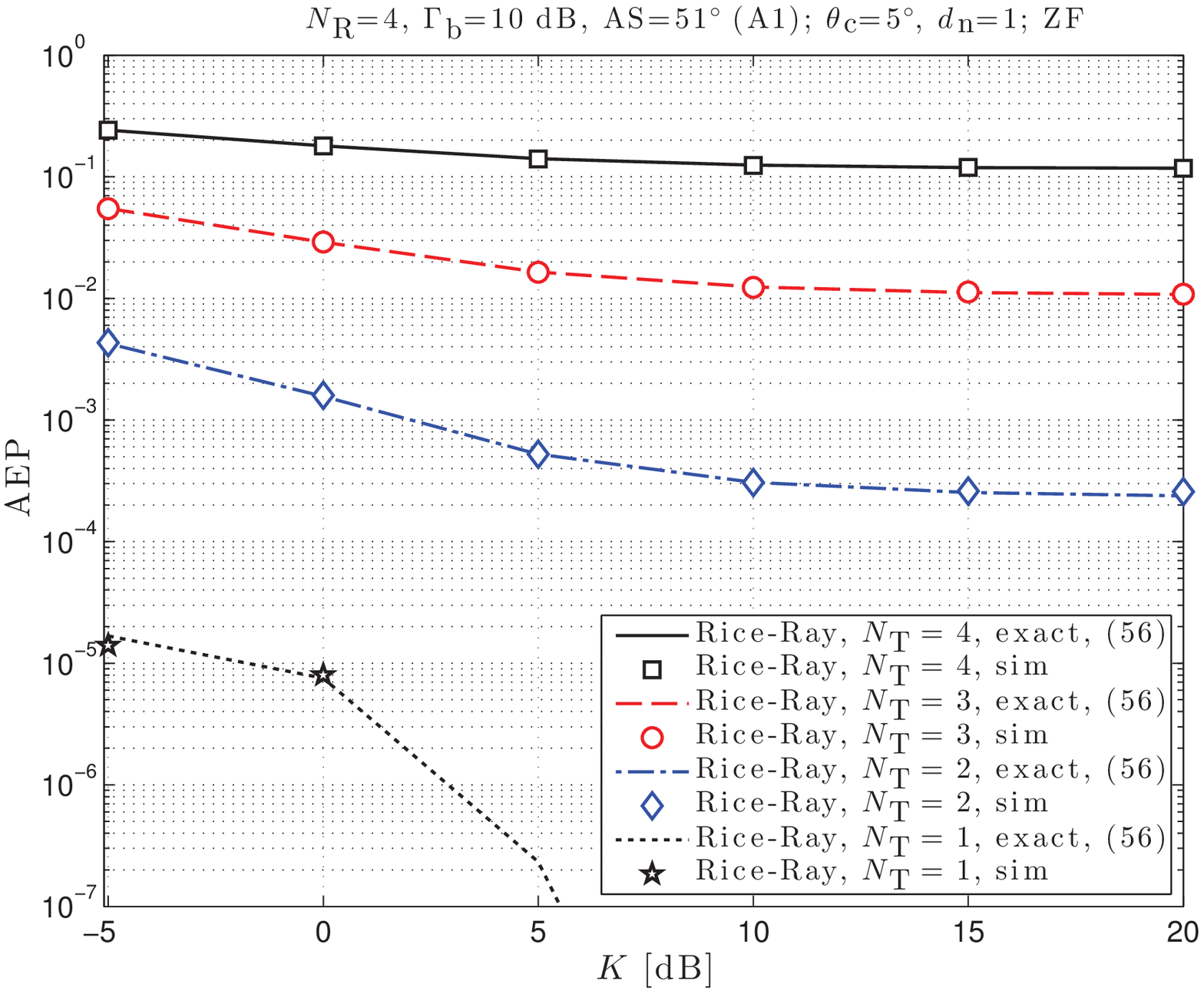}
\caption{Stream-1 AEP for $ \NR = 4 $, $ \NT = 1 : 4 $, $ \text{AS}= 51^\circ $, $  { \Gamma_{\text{b}} }  = 10 $~dB.}
\label{figure_AEP_AER_vs_K_ZF_Rice_Avg_AS_A1_NT_1_4_NR4}
\end{center}
\end{figure}

Fig.~\ref{figure_AEP_AER_vs_K_ZF_Rice_Avg_AS_A1_NT_1_4_NR4} shows that the AEP for the Rician-fading stream is decreasing with increasing $ K $ until it reaches a floor that increases dramatically with the number of Rayleigh-fading interfering streams, i.e., $ \NT - 1 $, which is  because the diversity order $ N =\NR - \NT + 1 $ decreases.
The facts that the AEP is independent of $ \ud_{\text{n}} $ and that it can remain high even for large $ K $ ($ \ud_{\text{n}} $ and $ K $ are the only likely controllable propagation features) are of concern for heterogeneous networks.
McKay \textsl{et al.} have observed similar issues for MIMO optimum combining, i.e., MMSE\cite[Fig.~3]{mckay_tcomm_09}.

\begin{figure}[t]
\begin{center}
\includegraphics[width=3.4in]
{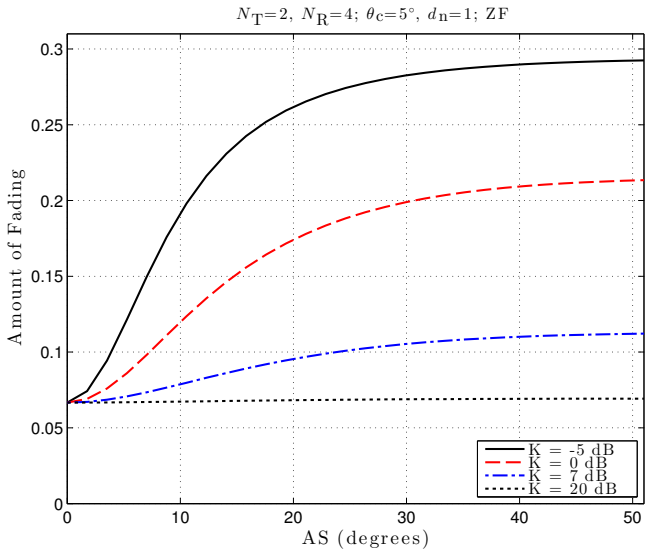}
\caption{Amount of fading corresponding to Stream 1, for $ \NR = 4 $, $ \NT = 2 $, and $ K = -5, 0, 7, 20 $ dB.}
\label{figure_Amount_Fading_ZF_vs_AS_K_NR_4_NT2_NR4}
\end{center}
\end{figure}

Fig.~\ref{figure_Amount_Fading_ZF_vs_AS_K_NR_4_NT2_NR4} shows, for $ \NT = 2 $, that the amount of fading is lower with higher $ K $, as expected.
Furthermore, for higher $ K $, the amount of fading varies less with the AS, which corroborates the observation on Fig.~\ref{figure_AEP_vs_SNR_ZF_Rice_Ray_A1_C2_D1_NT_2_NR_4} that AS does not affect performance  for large enough $ K $.

\begin{figure}[t]
\begin{center}
\includegraphics[width=3.4in]
{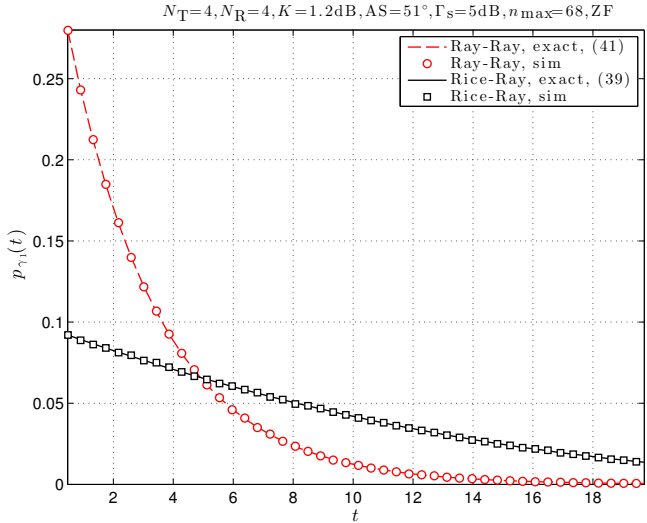}
\caption{P.d.f.~of the SNR (in linear units) for Stream 1, for $ \NR = 4 $, $ \NT = 4 $, $ K = 1.2 $ dB, $ \text{AS}= 51^\circ $, $  { \Gamma_{\text{s}} }  = 5 $~dB.}
\label{figure_ZF_SNR_pdf_Rayleigh_Rice_Fixed_AS_K_12_A1_NT4_NR4}
\end{center}
\end{figure}

\begin{figure}[t]
\begin{center}
\includegraphics[width=3.4in]
{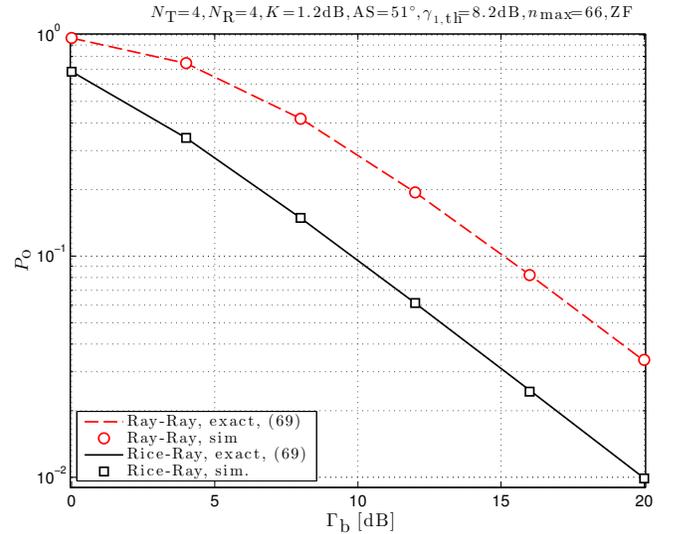}
\caption{Stream-1 outage probability for $ \NR = 4 $, $ \NT = 4 $, $ K = 1.2 $ dB, $ \text{AS}= 51^\circ $, and $ \gamma_{1,\text{th}} = 8.2 $~dB.}
\label{figure_ZF_SNR_Po_Rayleigh_Rice_Fixed_AS_K_12_A1_NT4_NR4}
\end{center}
\end{figure}

\begin{figure}[t]
\begin{center}
\includegraphics[width=3.4in]
{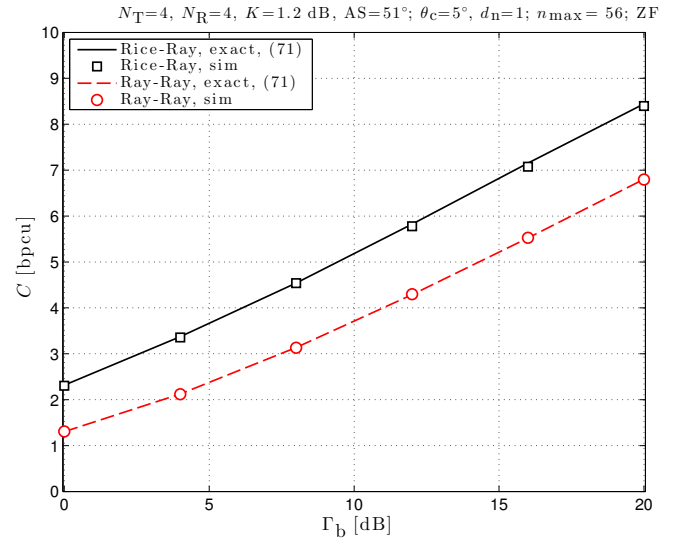}
\caption{Stream-1 ergodic capacity for $ \NR = 4 $, $ \NT = 4 $, $ K = 1.2 $~dB, $ \text{AS} = 51^{\circ} $.}
\label{figure_ZF_SNR_C_Rayleigh_Rice_Fixed_AS_K_12_A1_NT4_NR4}
\end{center}
\end{figure}

Fig.~\ref{figure_ZF_SNR_pdf_Rayleigh_Rice_Fixed_AS_K_12_A1_NT4_NR4} shows close agreement between the p.d.f.~of $ \gamma_1 $ (in linear units) from the new exact expression~(\ref{equation_gamma1_pdf_final}) and from simulation, for $ \NR = 4 $, $ \NT = 4 $, $ \text{AS}= 51^\circ $, $ K = 1.2 $ dB, and $ { \Gamma_{\text{s}} } = 5 $~dB.
These p.d.f.~plots also show that Rician fading for the intended stream tends to yield higher SNR values than Rayleigh fading.
Then, Fig.~\ref{figure_ZF_SNR_Po_Rayleigh_Rice_Fixed_AS_K_12_A1_NT4_NR4} shows close agreement between the outage probability from the new exact expression~(\ref{equation_gamma1_cdf_final}) and from simulation.
The threshold-SNR $ \gamma_{1,\text{th}} $ has been set to $ 8.2 $~dB, which corresponds for QPSK to the relevant error probability value $ P_{\text{e,th}} = 10^{-2} $\cite{siriteanu_tvt_09}.
These $ P_{\text{o}}  $ plots also indicate a diversity order of $ N $ for both Rician--Rayleigh and Rayleigh--Rayleigh fading, with the former displaying an array gain over the latter.
Further, Fig.~\ref{figure_ZF_SNR_C_Rayleigh_Rice_Fixed_AS_K_12_A1_NT4_NR4} shows close agreement\label{claim_good_agreement} between the ergodic capacity from the new exact expression~(\ref{equation_capacity_ergodic_inf_sum}) and from simulation.
Note that also these plots for Rician--Rayleigh and Rayleigh--Rayleigh fading are parallel at high $ \Gamma_{\text{b}} $, where the former outperforms the latter by about 1.5 bpcu.
Finally, the title in Fig.~\ref{figure_ZF_SNR_C_Rayleigh_Rice_Fixed_AS_K_12_A1_NT4_NR4} reveals that numerical convergence has occurred (in average over $ \Gamma_{\text{b}} $) for $ n = n_{\text{max}} = 56 $, i.e., just under the limit of numerical stability\footnote{Factorials of large numbers are represented inaccurately. E.g., in \texttt{MATLAB}, $ 50! \approx {\cal{O}}(10^{64} ) $ has representation error of $ {\cal{O}}(10^{48} )$. Error compounding yields numerical instability.} for the computation of $ B_n $ with~(\ref{equation_Bn})\cite[Sec.~V]{siriteanu_ausctw_14}.



\subsection{Description of Results for Rayleigh--Rician and Rayleigh--Rayleigh Fading}

\begin{figure}[t]
\begin{center}
\includegraphics[width=3.4in]
{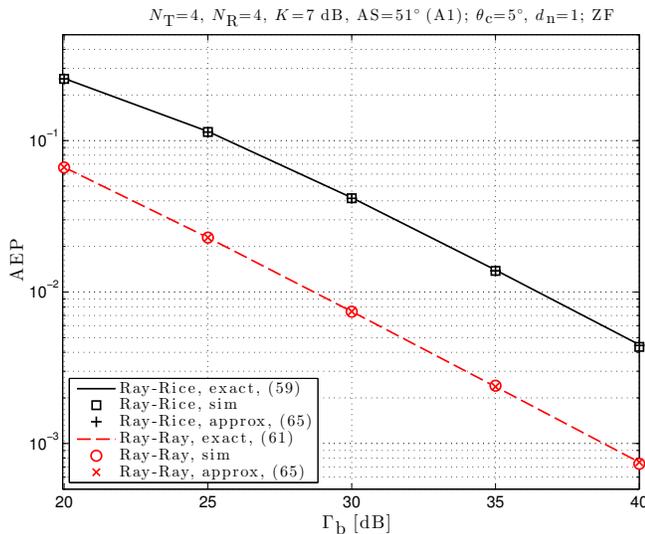}
\caption{Stream-1 AEP for $ \NR = \NT =4 $, $ K = 7 $~dB, $ \text{AS} = 51^{\circ} $.}
\label{figure_AEP_AER_vs_SNR_ZF_Rayleigh_Rice_Fixed_AS_K_A1_NT4_NR4}
\end{center}
\end{figure}

Fig.~\ref{figure_AEP_AER_vs_SNR_ZF_Rayleigh_Rice_Fixed_AS_K_A1_NT4_NR4} depicts, for $ \text{AS} = 51^{\circ} $, the AEP for Rayleigh--Rician fading, i.e., $ \uhda = \mzero $, $ \uHdb \neq \mzero $, and $ K = 7 $~dB, as well as for Rayleigh--Rayleigh fading.
The large AS value yields low transmit-correlation, so that $ \uRT \approx \mbI_{\NT} $, which implies $ \rTba \approx 0 $, i.e., $ \ur_{2,1} \approx \mzero $.
Thus, condition~(\ref{equation_hd1_Hd2_r_cond}) holds approximately, which explains the agreement revealed by the figure between the AEP from exact expression~(\ref{equation_aep_sum_Rayleigh_Rician_Cond}) and from simulation.
For low AS values, the AEP from~(\ref{equation_aep_sum_Rayleigh_Rician_Cond}) and from simulation no longer agree, which is because $ \uRT \not \approx \mbI_{\NT} $, so that $ \uhda \not \approx \uHdb \ur_{2,1} $.

The results for Rayleigh--Rician fading in Fig.~\ref{figure_AEP_AER_vs_SNR_ZF_Rayleigh_Rice_Fixed_AS_K_A1_NT4_NR4} are for $ \uHdb $ with equal elements.
Several other choices of $ \uHdb $ have yielded the same AEP results.
Thus, when intended Rayleigh fading is uncorrelated with the interfering Rician fading, the mean of the latter does not affect ZF performance, which is expected because the AEP from~(\ref{equation_aep_sum_Rayleigh_Rician_Cond}) is independent of $ \uHdb $.

Finally, Fig.~\ref{figure_AEP_AER_vs_SNR_ZF_Rayleigh_Rice_Fixed_AS_K_A1_NT4_NR4} reveals that the AEP from the exact expression~(\ref{equation_aep_sum_Rayleigh_Rician_Cond}) matches that from the approximate AEP expression~(\ref{equation_aep_approx_Ray}).
This surprising result is being investigated\cite{siriteanu_tit_13}.

\section{Summary and Conclusions}
We have derived exact infinite-series expressions for critical performance measures for zero-forcing detection of MIMO spatially multiplexed streams, in transmit-correlated fading that may be Rician either on the detected stream or on the interfering streams (but not on both).
Numerical results from our analysis agree with Monte Carlo simulations, and have offered new insights into effects of interference and channel fading statistics on ZF performance.
We have found ZF symbol-detection performance to be: 1) unaffected by the `direction' of the mean of the Rician-fading channel vectors; 2) largely unaffected by transmit-correlation, at realistic $K$ values; 3) dramatically degraded by more interferers, even for large $ K $, which is relevant for femtocells.

\section*{Acknowledgments}
\addcontentsline{toc}{section}{Acknowledgment}
This work was supported by: 1) Japan Science and Technology Agency, CREST; 2) Natural Sciences and Engineering Research Council of Canada  (NSERC) Discovery Grant 41731; 3) Grant KHU-20130436, 2013, Kyung Hee University, South Korea; Grant 2009-0083495 from National Research Foundation of Korea (NRF), funded by Ministry of Science, ICT and Future Planning; Grant NRF-2011-220-D00076 by Ministry of Education of Korea; 4) NSERC DAS project.

\footnotesize


\begin{IEEEbiography}[{\includegraphics[width=1in,height=1.25in,clip,keepaspectratio]{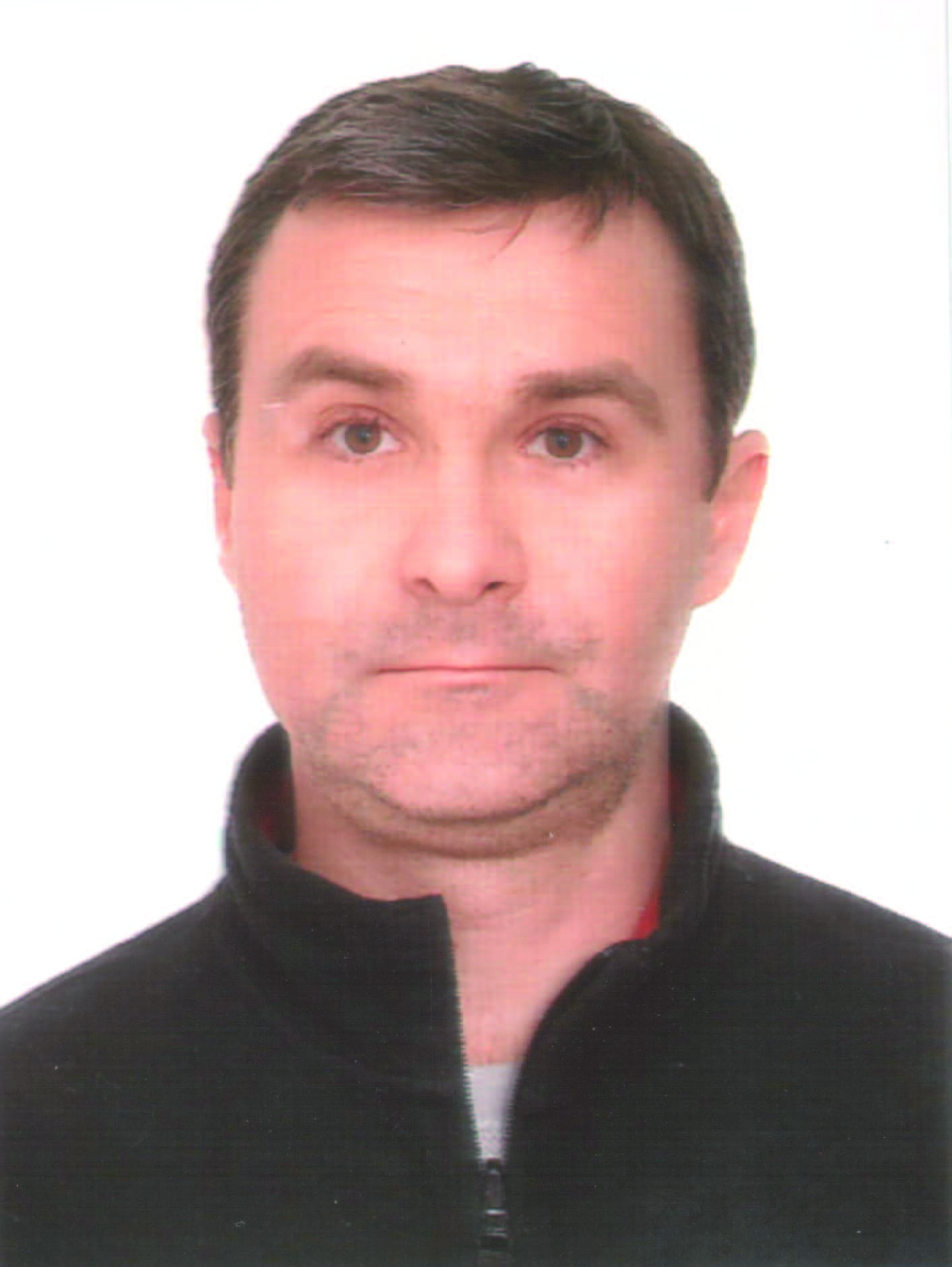}}]{Constantin (Costi) Siriteanu}
was born in
Sibiu, Romania. He received the Bachelor
and Master degrees in automatic control systems
from ``Gheorghe Asachi" Technical University, Iasi,
Romania, in 1995 and 1996, respectively, and the
Ph.D. degree from Queen's University, Canada, in 2006.
His Ph.D. thesis was on smart
antenna performance analysis and optimization.
Since 2006, he has done research and teaching at
universities in Korea, Canada, and Japan.
Since April 2013, he has been a Research Fellow with the Department of Mathematical Informatics, University of Tokyo.
His research interests are in developing multivariate statistics and probability concepts
that help analyze and evaluate the performance of multiple-input/multiple-output (MIMO) wireless
communications systems under realistic statistical assumptions about channel fading.
\end{IEEEbiography}

\begin{IEEEbiography}[{\includegraphics[width=1in,height=1.25in,clip,keepaspectratio]{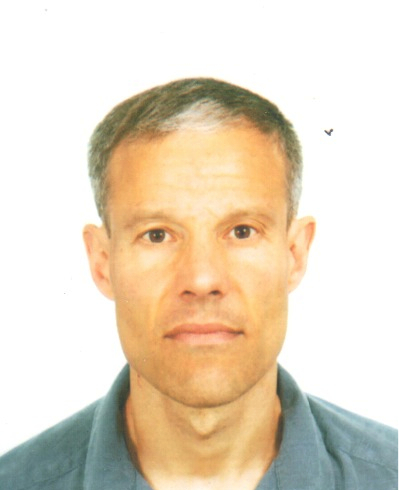}}]{Steven D. Blostein}
(SM '83, M '88, SM '96) received his B.S. degree in
Electrical Engineering from Cornell University, Ithaca, NY, in 1983,
and the M.S. and Ph.D. degrees in Electrical and Computer Engineering
from the University of Illinois, Urbana-Champaign, in 1985 and
1988, respectively. He has been on the faculty in the Department of
Electrical and Computer Engineering Queen's University since 1988 and
currently holds the position of Professor. From 2004-2009 he was Department Head.
He has also been a consultant to industry and government in the areas of
image compression, target tracking, radar imaging and wireless
communications. His current interests lie in the application of
signal processing to wireless communications systems,
including synchronization, cooperative and network MIMO, dynamic spectrum access, and cross-layer optimization for multimedia transmission.  He has been a member
of the Samsung 4G Wireless Forum and an invited distinguished
speaker.  He served as Chair of IEEE Kingston Section (1994), Chair of
the Biennial Symposium on Communications (2000, 2006, 2008), Associate
Editor for IEEE Transactions on Image Processing (1996-2000),
Publications Chair for IEEE ICASSP 2004, and Editor of IEEE
Transactions on Wireless Communications (2007-2013), and
served on numerous Technical Program Committees for IEEE Communications Society
conferences. He is a registered Professional Engineer
in Ontario and a Senior Member of IEEE.
\end{IEEEbiography}

\begin{IEEEbiography}[{\includegraphics[width=1in,height=1.25in,clip,keepaspectratio]{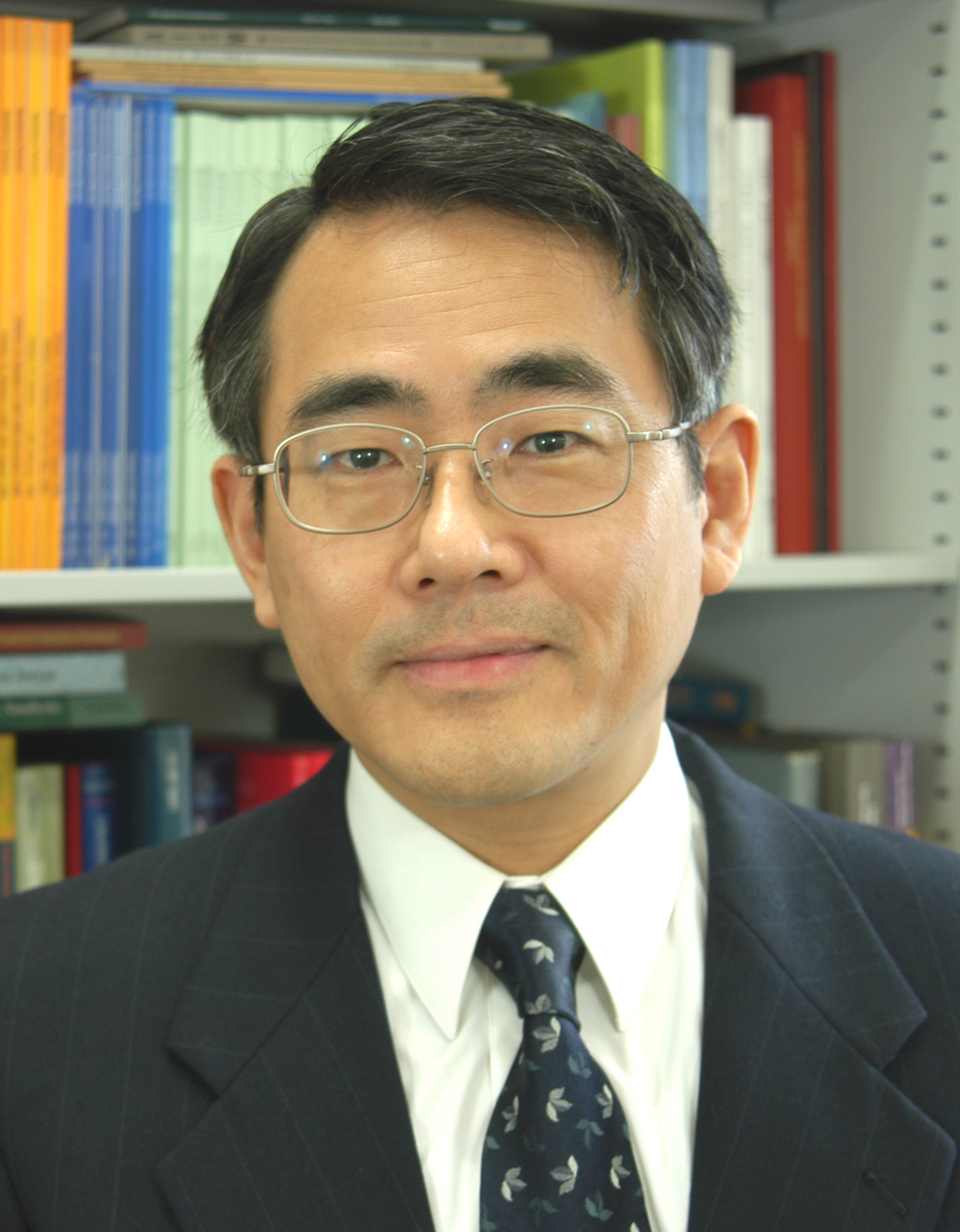}}]{Akimichi Takemura}
received the Bachelor of Arts degree in Economics in 1976 and
the Master of Arts degree in Statistics in 1978 from University of Tokyo, and
the Ph.D. degree in Statistics in 1982 from Stanford University.
He was an acting Assistant Professor at the Department of Statistics, Stanford University
from September 1992 to June 1983, and a
visiting Assistant Professor at the Department of Statistics, Purdue
University from  September 1983 to May 1984.
In June 1984 he has joined University of Tokyo, where he has been a Professor of Statistics with the Department of Mathematical Informatics since April 2001.
He has served as President of Japan Statistical Society from January 2011 to June 2013.
He has been working on multivariate distribution theory in statistics.
Currently his main area of research is algebraic statistics.  He also works on game-theoretic probability,
which is a new approach to probability theory.
\end{IEEEbiography}

\begin{IEEEbiography}[{\includegraphics[width=1in,height=1.25in,clip,keepaspectratio]{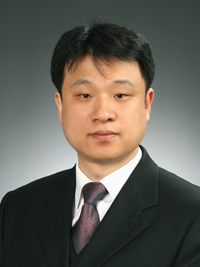}}]{Hyungdong Shin}(S’01-M’04-SM’11) received the B.S. degree in electronics engineering
from Kyung Hee University, Korea, in 1999, and the M.S. and Ph.D. degrees in electrical
engineering from Seoul National University, Korea, in 2001 and 2004, respectively. During
his postdoctoral research at the Massachusetts Institute of Technology (MIT) from 2004 to
2006, he was with the Wireless Communication and Network Sciences Laboratory within the
Laboratory for Information Decision Systems (LIDS).
In 2006, Dr. Shin joined Kyung Hee University, Korea, where he is now an Associate
Professor at the Department of Electronics and Radio Engineering. His research interests
include wireless communications and information theory with emphasis on MIMO
systems, cooperative and cognitive communications, network interference, vehicular
communication networks, location-aware radios and networks, physical-layer security, and
molecular communications.
Dr. Shin was honored with the Knowledge Creation Award in the field of Computer Science
from Korean Ministry of Education, Science and Technology (2010). He received the IEEE
Communications Society's Guglielmo Marconi Prize Paper Award (2008) and William R.
Bennett Prize Paper Award (2012). He was an Editor for IEEE Transactions on Wireless Communications
(2007-2012). He is currently an Editor for IEEE Communications Letters.
\end{IEEEbiography}

\begin{IEEEbiography}[{\includegraphics[width=1in,height=1.25in,clip,keepaspectratio]{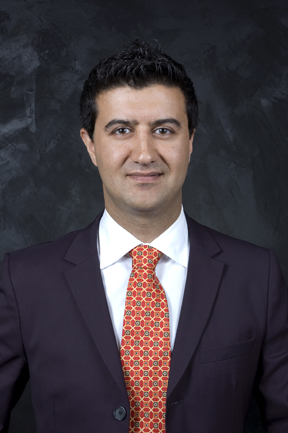}}]{Shahram Yousefi}
received his B.Sc. degree in Electrical
Engineering from University of Tehran, Iran, and his Ph.D. degree from the Department of Electrical and Computer
Engineering of University of Waterloo,
Canada, in September 2002. Since January 2003
he has been with the Department of Electrical
and Computer Engineering of Queen’s University,
Kingston, Canada, where he is a tenured Associate Professor. His research interests include wireless and wired networks,
signal design, MIMO and space-time systems, error control coding, and
network coding.
He currently serves as an Associate Editor for the
IEEE Communications Letters.
\end{IEEEbiography}

\begin{IEEEbiography}[{\includegraphics[width=1in,height=1.25in,clip,keepaspectratio]{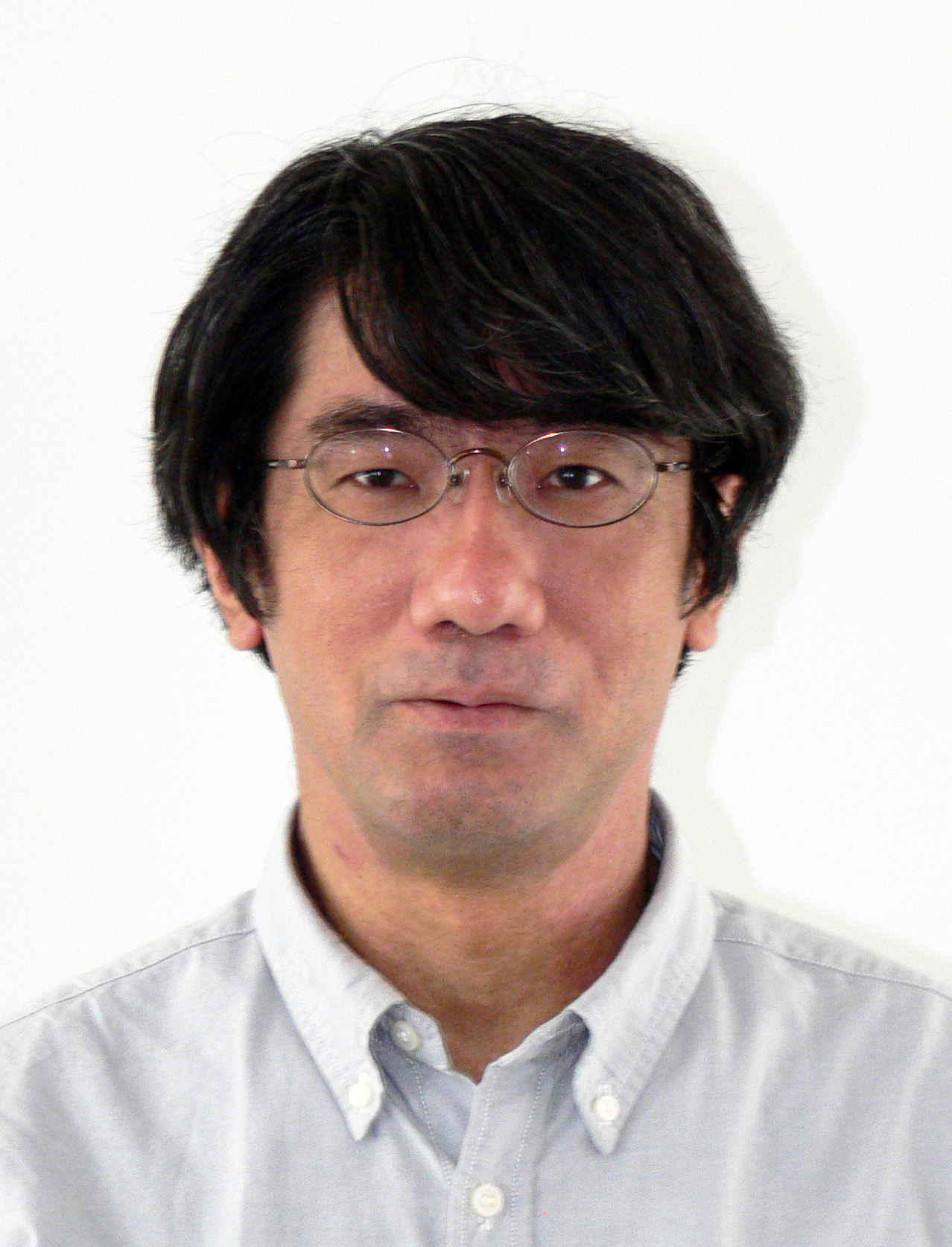}}]{Satoshi Kuriki}
received the Bachelor and Ph.D. degrees from University of Tokyo, Japan, in 1982 and 1993, respectively.  He is a Professor with the Institute of Statistical Mathematics (ISM), Tokyo, Japan, where he is also serving as Director of the Department of Mathematical Analysis and Statistical Inference.  His current major research interests include geometry of random fields, multivariate analysis, multiple comparisons, graphical models, optimal designs, and genetic statistics.
\end{IEEEbiography}

\end{document}